\documentclass[10pt]{article}
\usepackage[cp866]{inputenc}
\begin{document}

\centerline {{\Large\bf Evolutionary forms: Conservation laws and }}
\centerline {{\Large\bf causality}}
\centerline {L.I. Petrova}
\centerline{{\it Moscow State University, Russia, e-mail: ptr@cs.msu.su}}
\bigskip

Evolutionary forms are skew-symmetric differential forms the basis of which,
as opposed to exterior forms, are deforming manifolds (with unclosed metric
forms). Such differential forms arise when describing physical processes.

A specific feature of evolutionary forms is the fact that from the
evolutionary forms, which correspond to the conservation laws for
material media, the closed exterior forms, which correspond to the
conservation laws for physical fields, are obtained. This shows
that material media generate physical fields. And by this the
determinacy of physical processes and phenomena is revealed.

In this paper we obtain the mathematic apparatus that allows to
describe discrete transitions and quantum jumps. This relates to the fact that
the mathematic apparatus of exterior and evolutionary forms, which
basis involves nonidentical relations and degenerate
transformations, {\it can describe transitions from nonconjugate
operators to conjugate ones}. None of mathematic formalisms
contains such possibilities.

The physical results that disclose a mechanism of evolutionary processes
in material media and a generation of physical fields are obtained. These results
explain many actual processes.
\bigskip

{\large\bf Introduction}

In the paper a role of skew-symmetric differential forms in mathematic physics
and field theory is demonstrated.

It is known that the theories describing physical fields are based
on the invariant methods of investigation (tensor, group,
variational methods, theories of symmetries, transformations and
so on) derived from the postulates.

Mathematical physics, which describes physical processes in material media, is
based on the theory of differential equations.

In present paper it is shown that both mathematical physics and field theories
are a whole. This is obtained on the basis of exterior and evolutionary
skew-symmetric differential forms.

Closed exterior forms, which are differentials, have invariant properties.
The invariant properties of exterior forms explicitly or implicitly manifest
themselves essentially in all formalisms of field theory, such as the Hamilton
formalism, tensor approaches, group methods, quantum mechanics equations,
the Yang-Mills theory and others. They lie at the basis of field theory.

Evolutionary skew-symmetric differential forms are derived from differential
equations describing physical processes in material media and forming the basis
of mathematical physics.

In the paper it is shown that the closed exterior forms are obtained
from evolutionary forms. This firstly disclose a relation between mathematical
physics being based on differential equations and the invariant field theories.
And secondly, this proves that material media generate physical fields. Thus it
is disclosed a determinacy of physical processes and phenomena.

Such a role of exterior and evolutionary forms in physics relates to the fact
that exterior and evolutionary forms reflect the properties of conservation
laws, which, as it will be shown, play a controlling role in evolutionary
processes.

\bigskip
In essence, in present paper two problems are solved. 

Firstly, the
mathematical apparatus, which contains nontraditional elements,
namely, nonidentical relations and degenerate transformations, and
allows to describe evolutionary processes and discrete transitions, 
has been obtained. Non of mathematical
formalisms possesses such possibilities. Unique utilitarian
possibilities of mathematical apparatus of skew-symmetric
differential forms, which can serve as an approach to field
theory, are shown.

And secondly, a mechanism of evolutionary processes in material media, 
which leads to generation of physical fields, is described.

One can judge about the content of present paper by the following list.

{\it 1. Skew-symmetric differential forms.}

1.1 Some properties of manifolds. (The properties of manifolds are
connected with a difference between exterior and evolutionary differential forms.)

1.2 Specific features of skew-symmetric differential forms.

1.3 Closed exterior differential forms.
(Invariance, conjugacy, duality, and symmetries. Operators,
identical relations and nondegenerate transformations).

1.4 Evolutionary  skew-symmetric differential forms.
(Nonclosure. Nonidentical relations and degenerate transformations. Selfvariation
of nonidentical relation. Obtaining closed exterior forms from unclosed evolutionary
forms. Transition from nonidentical relation to identical one under degenerate
transformation. Relation between degenerate and nondegenerate transformations.
Mechanism of realization of conjugated objects and operators.)

1.5 Evolutionary forms: Realization of differential-geometrical 
structures and forming pseudometric and metric manifolds. 
(Characteristics and classification of differential-geometrical 
structures realized.) 

{\it 2. Role of exterior forms in field theory.}

2.1 Conservation laws for physical fields (exact conservation laws).

2.2 Closed exterior differential forms in invariant field theories.
(Exact conservation laws and specific features of existing field theories.) 

{\it 3. Role of evolutionary forms in mathematical physics and field theory.}

3.1 Balance conservation laws.

3.2 Evolutionary process in material medium and origination of physical
structures. (Nonequilibrium state of material system. Selfvariation of
nonequilibrium state of material system.
Transition of material system into locally equilibrium state.
Origination of physical structures.)

{\it 4. Evolutionary forms: Properties of physical structures.
Forming physical fields and manifolds.}

4.1 Characteristics of physical structures.

4.2 Relation between  physical structures originated and material systems.
(Potential forces.)

4.3 Characteristics of the formation created: intensity, spin,
absolute and relative speeds of propagation of the formation.

4.4 Forming pseudometric and metric spaces. (Space of gravitational field).

4.5 Forming physical fields. Classification of physical
structures.

{\it 5. Conservation laws. Symmetries. Causality.}

5.1 Conservation laws. (Relation between conservation laws for material
systems and conservation laws for physical fields.)

5.2 Symmetries.

5.3 Causality.

{\it 6. Certain aspects of quantum field theory and approaches to
general field theory.}

6.1 On interactions and classification of physical structures and physical fields.

6.2 Mathematical apparatus of skew-symmetric differential forms as the
basis of the evolutionary field theory. 

In {\it Appendix} presented at the end of the paper the analysis of
balance conservation laws for thermodynamic and gas dynamic systems and for 
the system of charged particles is given.

\section{Skew-symmetric differential forms}

\subsection{Some properties of manifolds.}

A distinction of evolutionary skew-symmetric differential forms
from exterior forms [1] is connected with the properties of manifolds
on which skew-symmetric forms are defined.

It is known that manifolds with structures of any type can serve as a
basis of exterior differential forms.
They have one common property, namely,
locally they admit one-to-one mapping into the Euclidean subspaces
and into other manifolds or submanifolds of the same dimension [2].

While describing the evolutionary processes in material systems
(material media)
one is forced to deal with manifolds which do not allow one-to-one
mapping described above. These can be manifolds constructed of trajectories
of material system elements (particles). Such manifolds, which can be called
accompanying manifolds, are deforming variable manifolds.
In real processes such varying deforming manifolds cannot be
manifolds of the types for which the theory of exterior
differential forms has been developed.

The differential forms defined on these manifolds are evolutionary ones.
The coefficients of these differential forms and the characteristics of
accompanying manifolds are interconnected and are varied
as functions of evolutionary variables.

A difference of manifolds on which exterior and evolutionary forms are
defined relates to the properties of metric forms of these manifolds.

Assume that on the manifold one can set the
coordinate system with base vectors $\mathbf{e}_\mu$ and define
the metric forms of manifold [3]: $(\mathbf{e}_\mu\mathbf{e}_\nu)$,
$(\mathbf{e}_\mu dx^\mu)$, $(d\mathbf{e}_\mu)$. The metric forms
and their commutators define the metric and differential
characteristics of the manifold.

If metric forms are closed
(the commutators are equal to zero), the metric is defined
$g_{\mu\nu}=(\mathbf{e}_\mu\mathbf{e}_\nu)$, and the results of
translation over manifold of the point
$d\mathbf{M}=(\mathbf{e}_\mu dx^\mu)$ and of the unit frame
$d\mathbf{A}=(d\mathbf{e}_\mu)$ prove to be independent of the
curve shape (the path of integration).

To describe the manifold differential characteristics
and, correspondingly, the metric form commutators, one can use
connectednesses [2-4].
If the components of metric form can be
expressed in terms of connectedness $\Gamma^\rho_{\mu\nu}$ [3],
the expressions $\Gamma^\rho_{\mu\nu}$,
$(\Gamma^\rho_{\mu\nu}-\Gamma^\rho_{\nu\mu})$ and
$R^\mu_{\nu\rho\sigma}$ are components of the commutators of
the metric forms with zeroth- first- and third degrees. (The commutator of
the second degree metric form is written down in a more complex
manner [3], and therefore it is not presented here).

The closed metric forms define the manifold structure. 
And the commutators of metric forms
define the manifold differential characteristics that specify
the manifold deformation: bending, torsion, rotation, and twist.
(For example, the commutator of the zeroth degree metric form
$\Gamma^\rho_{\mu\nu}$ characterizes the bend, that of the first degree
form $(\Gamma^\rho_{\mu\nu}-\Gamma^\rho_{\nu\mu})$ characterizes the torsion,
the commutator of the third -degree metric form $R^\mu_{\nu\rho\sigma}$
determines the curvature. (For manifolds with closed metric form of first
degree the coefficients of connectedness are symmetric ones.)

Is is evident that the manifolds that are metric ones or possess the
structure have closed metric forms. It is with such manifolds that the
exterior differential forms are connected.

If the manifolds are deforming manifolds, this means that their
metric form commutators are nonzero. That is, the metric forms of such
manifolds turn out to be unclosed.
The accompanying manifolds appearing to be deforming ones are the
examples of such manifolds.

The skew-symmetric evolutionary differential forms
whose basis are accompanying deforming manifolds are defined
on manifolds with unclosed metric forms.

Below it will be shown that the specific properties of evolutionary skew-symmetric
differential forms, which are defined on manifolds with unclosed metric forms,
are connected with the properties of metric form commutators.

\subsection{Specific features of skew-symmetric differential forms}

The skew-symmetric differential form of degree $p$ ($p$-form) can be
written down as [4,5] 
$$
\omega^p=\sum_{\alpha_1\dots\alpha_p}a_{\alpha_1\dots\alpha_p}dx^{\alpha_1}\wedge
dx^{\alpha_2}\wedge\dots \wedge dx^{\alpha_p}\quad 0\leq p\leq n\eqno(1.1)
$$
where the local basis obeys the condition
$$
\begin{array}{l}
dx^{\alpha}\wedge dx^{\alpha}=0\\
dx^{\alpha}\wedge dx^{\beta}=-dx^{\beta}\wedge dx^{\alpha}\quad
\alpha\ne \beta
\end{array}
$$

The differential of skew-symmetric differential form can be written in the form
$$
d\omega^p{=}\!\sum_{\alpha_1\dots\alpha_p}\!da_{\alpha_1\dots\alpha_p}dx^{\alpha_1}dx^{\alpha_2}\dots
dx^{\alpha_p}{+}\!\sum_{\alpha_1\dots\alpha_p}\!a_{\alpha_1\dots\alpha_p}d(dx^{\alpha_1}dx^{\alpha_2}\dots
dx^{\alpha_p})\eqno(1.2)
$$
where the second term is connected with the differential of the basis.

[In further presentation a symbol of summing $\sum$ and a symbol
of exterior multiplication $\wedge$ will be omitted. Summation
over repeated indices is implied.]

The second term connected with the differential of the basis is expressed in
terms of the metric form commutator.
For differential forms defined on manifold with unclosed metric
form one has $d(dx^{\alpha_1}dx^{\alpha_2}\dots
dx^{\alpha_p})\neq 0$.
And for manifold with closed metric form it is valid the following
$d(dx^{\alpha_1}dx^{\alpha_2}\dots dx^{\alpha_p}) = 0$.

That is, for differential forms defined on the manifold with unclosed metric
form the second term is nonzero, whereas for differential forms
defined on the manifold with closed metric form the second term vanishes.

For example, let us consider the first-degree form
$\omega=a_\alpha dx^\alpha$. The differential of this form can
be written as
$$d\omega=K_{\alpha\beta}dx^\alpha dx^\beta\eqno(1.2')$$
where
$K_{\alpha\beta}=a_{\beta;\alpha}-a_{\alpha;\beta}$ are
the components of the commutator of the form $\omega$, and
$a_{\beta;\alpha}$, $a_{\alpha;\beta}$ are the covariant
derivatives. If we express the covariant derivatives in terms of
the connectedness (if it is possible), they can be written
as $a_{\beta;\alpha}=\partial a_\beta/\partial
x^\alpha+\Gamma^\sigma_{\beta\alpha}a_\sigma$, where the first
term results from differentiating the form coefficients, and the
second term results from differentiating the basis. (In the
Euclidean space the covariant derivatives coincide with ordinary ones
since in this case the derivatives of the basis vanish). If
we substitute the expressions for covariant derivatives into the
formula for commutator components, we obtain the following expression
for the commutator components of the form $\omega$
$$
K_{\alpha\beta}=\left(\frac{\partial a_\beta}{\partial
x^\alpha}-\frac{\partial a_\alpha}{\partial
x^\beta}\right)+(\Gamma^\sigma_{\beta\alpha}-
\Gamma^\sigma_{\alpha\beta})a_\sigma\eqno(1.3)
$$
Here the expressions
$(\Gamma^\sigma_{\beta\alpha}-\Gamma^\sigma_{\alpha\beta})$
entered into the second term are just the components of
commutator of the first-degree metric form.

That is, the corresponding
metric form commutator will enter into the differential form commutator.

If we substitute the expressions (1.3) for the skew-symmetric differential form
commutator into formula (1.2'), we obtain the
following expression for the differential of the first degree skew-symmetric form
$$
d\omega=\left(\frac{\partial a_\beta}{\partial
x^\alpha}-\frac{\partial a_\alpha}{\partial
x^\beta}\right)dx^\alpha dx^\beta+\left((\Gamma^\sigma_{\beta\alpha}-
\Gamma^\sigma_{\alpha\beta})a_\sigma\right)dx^\alpha dx^\beta
$$
The second term in the expression for the differential of skew-symmetric form
is connected with the differential of the manifold metric form, which is expressed
in terms of the metric form commutator.

While deriving formula (1.3) for the differential form commutator
the connectednesses of special type were used. However, a similar
result can be obtained by applying the connectednesses of arbitrary type
or by using another way for finding the differential of the base coordinates.
For differential forms of any degree the metric form commutator of
corresponding degree will be included into the commutator of the skew-symmetric
differential form.

As it is known [4,5], the differential of exterior differential form
involves only a single term.
There is no second term. This indicates that the metric form commutator
vanishes. In other words, the manifold, on which the {\it exterior}
differential form is defined, has a closed metric form.

The differential of the evolutionary differential form, which is defined
on the manifold with unclosed metric forms, will contain two terms: the
first term depends on the differential form coefficients
and the other depends on the differential characteristics of the
manifold.

Thus, the differentials and, correspondingly, the commutators of exterior and
evolutionary forms are of different types.

What does this lead to?

The commutator of {\it exterior} differential form contains only
one term obtained from the derivatives of the differential form
coefficients. Such a commutator may be equal to zero or nonzero
depending on the form coefficients. This means that the exterior
differential form may be closed or unclosed.

In contrast to this, the evolutionary differential form cannot be
closed. The terms of the commutator of evolutionary differential
form have a different nature. Such terms cannot compensate one
another. For this reason, the evolutionary form commutator, and
hence, the differential of evolutionary form, prove to be nonzero. And
this means that the evolutionary form cannot be closed.

\subsection{Closed exterior differential forms}

\subsection*{Functional and utilitarian properties of closed exterior
differential forms}

Exterior differential form is called a closed one if its differential
is equal to zero:
$$
d\theta^p=0\eqno(1.4)
$$
From condition (1.4) one can see that the closed form is a conservative
quantity. This means that it can correspond to the conservation law, namely,
to some conservative physical quantity.

The differential of exterior form is a closed form.
The exterior form which is the differential of some other form
$$
\theta^p=d\theta^{p-1}\eqno(1.5)
$$
is called an exact form. (The exact forms prove to be closed
automatically).

The closed inexact exterior forms possess utilitarian properties.
These are exterior forms that are closed only on a certain
structure (with the dimension being less than that of the initial
manifold). In its metric form such a structure turns out to be a
pseudostructure.

If the exterior form is closed only on pseudostructure, the closure
condition is written as
$$
d_\pi\theta^p=0\eqno(1.6)
$$
And the pseudostructure $\pi$ obeys the condition
$$
d_\pi{}^*\theta^p=0\eqno(1.7)
$$
where ${}^*\theta^p$ is the dual form.

From conditions (1.6) and
(1.7) one can see that the pseudostructure and the closed exterior form
constitute a conservative object, namely, a quantity that is
conservative on the pseudostructure. Hence, such an object can
correspond to some conservation law. It is from such object that the
physical fields are formed. (More detail
description will be given below).

The exact form is, by definition, a differential.
In this case the differential is a total one.
The closed inexact form is a differential too. And in this case the
differential is an interior (on pseudostructure) differential, that is
$$
\theta^p_\pi=d_\pi\theta^{p-1}\eqno(1.8)
$$

Thus any closed form is a differential of the exterior form of a lower
degree. From this it follows that the exterior form of lower
degree may correspond to the potential, and the closed form by itself
may correspond to the potential force. This is an additional example
showing that the closed exterior form may have a physical sense. Here
the two-fold nature of the closed form is revealed, on the one hand,
as a conservative locally constant object, and, on the other hand, as
a potential force.

Similarly to the differential connection between the exterior
forms of sequential degrees, there is an integral connection.
In particular, the integral theorems by Stokes and Gauss follow
from the integral relation for $p=1,2$ in three-dimensional space.

\bigskip
{\bf Invariant properties of closed exterior differential
forms.}
Since the closed form is a differential (a total one if the form is exact,
or an interior one on the pseudostructure if the form is inexact), it is
obvious that the closed form turns out to be invariant under all
transformations that conserve the differential. The unitary transformations
(0-form), the tangent and canonical transformations (1-form), the gradient and
gauge transformations (2-form) and so on are  examples of such transformations.
{\it These are gauge transformations for spinor, scalar, vector, tensor
(3-form) fields}. It should be pointed out
that just such transformations are used in field theory.

\bigskip
{\bf Conjugacy and duality of exterior differential forms. Symmetries.}
The closure of exterior differential form, which leads to invariance, is a
result of the conjugacy of elements of exterior or dual forms.
The closure property of the exterior form means that any objects,
namely, elements of the exterior form, components of elements, elements
of the form differential, exterior and dual forms and others, turn
out to be conjugated. It is a conjugacy that leads to realization of
invariant and covariant properties of the exterior and dual
forms.

Conjugacy is possible if there is one or other type of symmetry. Gauge
symmetries, which are interior symmetries of field theory and with which
gauge transformations are connected, are the symmetries of closed exterior
differential forms. They are obtained as the result of conjugacy of any
exterior form elements. The conservation laws {\it for physical fields }
are connected with such interior symmetries.

(Below it will be shown that the exterior symmetries are symmetries
of evolutionary differential forms, which are conditioned by degrees
of freedom of {\it material systems}).

With the conjugacy another characteristic
property of the exterior differential forms, namely, their duality,
is connected. \{The conjugacy is some identical connection between
two operators or mathematical objects.  The duality is a
concept that means that one object carries a double meaning or
that two objects with different meanings (of different physical
nature) are identically connected. If one knows any dual object, one
can obtain the other object\}.

The connection of the exterior and dual forms is an example of the
duality. The exterior form and the dual form correspond to the
objects of different nature: the exterior form corresponds to the
physical (i.e. algebraic) quantity, and the dual form corresponds to some
spatial (or pseudospatial) structure. At the same time, under
conjugacy the duality of these objects manifests itself, that is,
if one form is known, it is possible to find the other form.
In the case of closed inexact form the relation between the closed
exterior form and the dual one leads to forming a differential-geometrical
structure. (Physical structures that form physical fields
are examples of such structures).

The duality is also revealed in the fact that, if the degree of the
exterior form equals $p$, the dimension of the structure
equals $N-p$, where $N$ is the space dimension.

As the closed exterior form possesses invariant properties and the dual form
corresponding to that possesses covariant properties, the
invariance and covariance is one of examples of duality of the exterior
differential forms.

The further example of the duality of the closed form is connected
with the fact that, on the one hand, the closed exterior form
is conservative quantity, and, on the other hand, the closed
form can correspond to potential force. (Below the physical
meaning of this duality will be elucidated, and it will be shown in
respect to what the closed form manifests itself as potential
force and with what the conservative physical quantity is
connected).

As it will be shown below, the duality of closed differential forms
has a fundamental physical meaning. The duality is a tool, which
untangles the mutual connection, the mutual changeability and the transitions
between different physical objects in evolutionary processes.

\subsection*{Specific features of the mathematical apparatus of exterior
differential forms.}

{\bf Operators of the theory of exterior differential forms.}
In  differential calculus the derivatives are basic elements of
the mathematical apparatus. By contrast, the differential is an element
of mathematical apparatus of the
theory of exterior differential forms. It enables one to analyze
the conjugacy of derivatives in various directions, which extends
the potentialities of differential calculus.

The operator of exterior differential $d$ 
is an abstract generalization of ordinary mathematical operations
of the gradient, curl, and divergence in the vector calculus [5].
[Here it should be emphasized that usual concepts of gradient,
curl and divergence are operators applied to vector, whereas a
gradient, curl and divergence in the theory of exterior forms,
which are obtained as a result of exterior differentiation, are
operators applied to pseudovector (an axial vector).]

If, in addition to the exterior differential, we
introduce the following operators: (1) $\delta$ for transformations that convert
the form of $p+1$ degree into the form of $p$ degree, (2) $\delta'$
for cotangent transformations, (3) $\Delta$ for the
$d\delta-\delta d$ transformation, (4)$\Delta'$ for the $d\delta'-\delta'd$
transformation, one can write down the operators in the field
theory equations in terms of these operators that act on the exterior
differential forms. The operator $\delta$ corresponds to Green's
operator, $\delta'$ does to the canonical transformation operator,
$\Delta$ does to the d'Alembert operator in 4-dimensional space, and
$\Delta'$ corresponds to the Laplace operator [6]. It can be seen
that the operators of the exterior differential form theory are
connected with many operators of mathematical physics.

The mathematical apparatus of exterior differential forms extends
the potentialities of the integral calculus. As it has been pointed out,
the exterior differential forms were introduced as integrand
expressions for definition of the integral invariants. The closure
condition for exterior differential forms makes it possible to find the
integrability conditions. This fact is of great importance
in applications.

\bigskip
{\bf Identical relations of exterior differential forms.}
The basis of the mathematical apparatus of closed exterior differential
forms is formed by identical relations. (Below it will be shown that at the
basis of the mathematical apparatus of evolutionary differential forms there
lie nonidentical relations, and it will be shown that identical
relations for exterior differential forms are obtained from nonidentical
relations for evolutionary forms.)

The identical relations of exterior differential forms are an
mathematical expression of various kinds of the conjugacy or the duality of
closed differential forms. (Since the conjugacy is a certain connection between two operators or
mathematical objects, it is evident that the relations can be used to
express the conjugacy mathematically.)
They describe the conjugacy of any objects: the form elements,
components of each element, exterior and dual forms, exterior forms of
various degrees, and others. The identical relations that are connected
with different kinds of conjugacy elucidate invariant, structural and group properties
of exterior forms, which is of great importance in applications.

The identical relations for exterior differential forms follow from
the closure conditions of differential forms, namely, vanishing the form
differential (see formulae (1.4), (1.6), (1.7)) and the conditions
connecting the forms of consequent degrees (see formulae (1.5), (1.8)).

The identical relation, which connects exterior forms of consequent degrees,
express the fact that each closed exterior form is a differential of some exterior
form. In general form such an identical relation can be written as
$$
d _{\pi}\phi=\theta _{\pi}^p\eqno(1.9)
$$
In this relation the form in the right-hand side has to be a {\it closed }
one. (As it will be shown below,
the identical relations are satisfied only on pseudostructures).

In identical relation (1.9) in one side it stands a closed form and
in other side does the  differential 
of some differential form (the differential is a closed form as well).
The examples of such identical relations are
a) the Poincare invariant $ds\,=-\,H\,dt\,+\,p_j\,dq_j$,
b) the second principle of thermodynamics $dS\,=\,(dE+p\,dV)/T$,
d) the conditions on characteristics in the theory of differential
equations and so on. 
The requirement that the function is an antiderivative (the integrand is a
differential
of a certain function) can be written in terms of such an identical relation.
Existence of harmonic function is
written by means of the identical relation: the harmonic function is a closed
form, that is, a differential. 

In addition to relations in differential forms, from
the closure conditions of differential forms and the conditions
connecting the forms of consequent degrees the identical relations of
other types are obtained. This is also connected with the fact that
the exterior forms can have the tensor, differential or integral representation.
As the example of such relations it can serve

{\it integral identical relations} (formulae by Newton, Leibnitz and Green and
integral relations by Stokes, Gauss-Ostrogradskii, etc);

{\it tensor identical relations} 
(vector and tensor identical relations that connect
the operators of gradient, curl, divergence and so on];

{\it identical relations between derivatives}
(the Cauchi-Riemann conditions in the theory of complex
variables, the transversality condition in the calculus of variations,
the canonical relations in the Hamilton formalism, the thermodynamic relations
between derivatives of thermodynamic functions [7], the condition that the
derivative of implicit function is subject to, the eikonal relations [8] and so on.

As the examples of identical relations it can serve such relations as the gauge
relations in electromagnetic field theory, the tensor relations between
connectednesses and their derivatives in gravitation (the symmetry of connectednesses 
with respect to subscripts,
the Bianchi identity, the conditions imposed on the Christoffel symbols) and so on.

The identical relations of exterior differential forms appear practically in
all branches of physics, mechanics and thermodynamics.

The physical meaning of identical relations for exterior differential forms
will be disclosed below using evolutionary differential forms.

\bigskip
{\bf Nondegenerate transformations.}
One of the fundamental methods in the theory of exterior
differential forms is an application of {\it nondegenerate} transformations.
The nondegenerate transformations, if applied to identical relations, enable one
to obtain new identical relations and new closed exterior differential
forms.

Nondegenerate transformations in the theory of exterior
differential forms are those that conserve the differential.
As it has been pointed, the unitary, tangent, canonical,
gradient, and gauge transformations are examples of such transformations.

From description of operators of exterior differential forms one can see that
they are operators which execute some transformations. All these transformations
are connected with the nondegenerate transformations of exterior
differential forms.

The significance of the nondegenerate transformations consists in the fact
that they allow to find closed differential forms.

\bigskip
As one can see, the properties of closed exterior differential forms
reveal in many branches of physics.

In the second part of the paper it will be demonstrated a role of exterior
differential forms in field theory. This role is connected with the fact
that closed exterior forms reflect the properties of conservation laws for
physical fields.

\subsection{Evolutionary skew-symmetric differential forms}

The properties of exterior differential forms and specific features of their
mathematical apparatus have a great utilitarian and functional
importance. However, the question of how closed exterior differential
forms are obtained and how the process of conjugating different objects
is realized remains unsolved.

The mathematical apparatus of evolutionary differential forms enables
us to answer this question.

\subsection*{Nonclosure of evolutionary differential forms.}

It has been shown above that the exterior skew-symmetric
differential forms can be closed. Such skew-symmetric differential
forms possess invariant properties, and the basis of their mathematical
apparatus includes identical relations and degenerate transformations. (Just
on these properties the field theory is based.)

The evolutionary forms, as opposed to exterior forms, cannot be closed.
This relates to the fact that the manifolds, on which the evolutionary forms
are defined, have unclosed metric forms. The commutator of unclosed metric 
form, which is nonzero, enters into the evolutionary form commutator. 
Such commutator of
differential form cannot vanish. This means that the evolutionary form
differential is nonzero. Hence, the evolutionary differential form,
in contrast to the case of the exterior form, cannot be closed and does not
possess invariant properties.

As it will be shown below, the basis of mathematical apparatus of evolutionary
differential forms, which are unclosed forms, includes {\it nonidentical}
relations and {\it degenerate } transformations. (Nonidentical relations under
degenerate transformations are just the relations that generate identical relations
for exterior forms.)

\subsection*{Nonidentical relations of evolutionary differential forms}

The identical relations of closed exterior differential forms reflect
a conjugacy or duality of any objects. The evolutionary forms,
being unclosed, cannot directly describe the conjugacy of any objects.
But they allow a description of {\it the process in which the conjugacy may appear}
(that is, to describe how closed exterior differential forms are generated).
Such a process is described by nonidentical relations.

The identical relations establish exact correspondence between the quantities
(or objects) involved into this relation. It is possible in the case
when the quantities involved into the relation are measurable (invariant)
ones.
In the nonidentical relations one of the quantities is unmeasurable.

[The concept of ``nonidentical relation"  may appear to be inconsistent.
However, as it will be shown below, this has a deep meaning.]

The relation of evolutionary forms is a relation between the differential,
which is a closed differential form and is a measurable (invariant)
quantity, and the evolutionary form that is unclosed form and hence is not
a measurable quantity. Such a relation cannot be identical.

Nonidentical relations of such type appear in descriptions of
physical processes.

Nonidentical relations may be written as
$$
d\psi \,=\,\omega^p \eqno(1.10)
$$
Here $\omega^p$ is the $p$-degree evolutionary form, which is unclosed form,
$\psi$ is a certain form of degree $(p-1)$, and
the differential $d\psi$ is a closed form of degree $p$.

In the left-hand side of this relation it stands the form differential,
i.e. a closed form that is an invariant object. In the right-hand
side it stands the unclosed form that is not an invariant object.
Such a relation cannot be identical. \{Nonidentical relation 
was analyzed in paper J.L.Synge "Tensorial Metods in Dynamics" (1936). 
And yet it was allowed a possibility to use the sign of 
equality in nonidentical relation.\}   
 
One can see a difference of relations for exterior forms and evolutionary ones.
In the right-hand side of identical relation (1.9)
it stands a closed form, whereas the form in the right-hand side of
nonidentical relation (1.10) is an unclosed one.

Relation (1.10) is an evolutionary relation as it involves the
evolutionary form.

Such nonidentical relations appear, for example, while
investigating the integrability of any differential equations that
describe various processes.

As the example one can inspect the partial differential equation of
the first order.
Let
$$ F(x^i,\,u,\,p_i)=0,\quad p_i\,=\,\partial u/\partial x^i \eqno(1.11)$$
be a partial differential equation of the first order.

Let us consider the functional relation
$$ du\,=\,\theta\eqno(1.12)$$
where $\theta\,=\,p_i\,dx^i$ Here $\theta\,=\,p_i\,dx^i$ is the
differential form of the first degree (summation over repeated
indices is implied).

In the general case, for example, when differential equation (1.11) describes any
physical processes, functional relation (1.12) turns out to be nonidentical.

For this relation be identical, the
differential form $\theta\,=\,p_i\,dx^i$ must also  be a differential
(like the left-hand side of relation (1.12)), that is, it has to be a
closed exterior differential form. To do this, it requires the commutator
$K_{ij}=\partial p_j/\partial x^i-\partial p_i/\partial x^j$ of the
differential form $\theta $ has to vanish.

In the general case, from equation (1.11) it does not follow (explicitly)
that the derivatives $p_i\,=\,\partial u/\partial x^i $, which obey
to the equation (and given boundary or initial conditions of the
problem), make up a differential. Without any supplementary conditions
the commutator $K_{ij}$ of the differential form $\theta $
is not equal to zero. The form $\theta\,=\,p_i\,dx^i$ turns out to be
unclosed and is not a differential like the left-hand side of relation
(1.12). Functional relation (1.12) appears to be nonidentical.

If to write functional relation (1.12) in the form
$$ du\,-\,p_i\,dx^i=0\eqno(1.12')$$
we get the well-known Pfaff equation for partial differential
equation. However, the nonidentical relation cannot be treated as
an equation in differentials. To solve  equation (1.12') means to
find the derivatives $p_i$ of original equation (1.11) which make up
a differential. In this case the derivatives $p_i$ of equation (1.11)
that do not obey these conditions are ignored, although they
satisfy original equation (1.11) and boundary and initial conditions.
But in the analysis of relation (1.12) all derivatives that satisfy
original equation (1.11) and boundary or initial conditions are
accounted for, and their role in the physical process under
consideration is investigated.

The nonidentity of functional relation means that the original equation is
nonintegrable one, that is, the derivatives of this equation are not reduced
(without additional condition) to the identical relation of the type $d\psi=dU$.

It can be shown that, for every differential equation describing any physical
processes, the relevant functional relation constructed of derivatives will be
nonidentical ones.

Below a role of such relations in mathematical physics will be disclosed.

While investigating real physical processes one often meets
relations that are nonidentical. Usually one treats nonidentical
relations like equations, namely, one finds only the values at
which nonidentical relation becomes identic. In doing so, it is
not accounted for the fact that the nonidentical relation is often
obtained from the description of some physical process and it has
physical meaning at every stage of the physical process rather
than at the stage when the additional conditions are satisfied.

The approach to nonidentical relation as to a relation rather then to the
equation in differentials gives fundamentally new physical results.

\bigskip
{\bf Selfvariation of nonidentical relation.}
Nonidentical relation possesses an unique evolutionary property, namely,
this relation appears to be a{\it selfvarying } relation.

The nonidentical relation is selfvarying one since, firstly, it is nonidentical,
namely, it contains two objects one of which appears to be unmeasurable 
(noninvariant), and,
secondly, it is an evolutionary relation, namely, the variation of
any object of the relation in some process leads to the variation of
another object, and, in turn, the variation of the latter leads to
variation of the former. Since one of the objects is an unmeasurable
quantity, the other cannot be compared with the first one, and hence,
the process of mutual variation cannot terminate.

Varying the evolutionary form coefficients leads to varying the first
term of the commutator (see Eq.(1.3) for the case of the first-order
evolutionary form). In accordance with this variation it varies the second term,
that is, the metric form of the manifold varies. Since the metric form
commutators specifies the manifold differential characteristics that are
connected with the manifold deformation, this points to the manifold
deformation.
This means that the evolutionary form basis varies. In turn, this leads to
variation of the evolutionary form, and the process of intervariation of the
evolutionary form and the basis is repeated. The processes of variation
of the evolutionary form and the basis are governed by the evolutionary
form commutator and it is realized according to the evolutionary relation.

Selfvariation of the evolutionary relation goes on by exchange between
the evolutionary form coefficients and the manifold characteristics.
This is an exchange between physical quantities and space-time
characteristics. (This is an exchange between quantities of different
nature).

The process of the evolutionary relation selfvariation cannot come to the end.
This is indicated by the fact that both the evolutionary form
commutator and the evolutionary relation involve unmeasurable
quantities. From this it follows that the evolutionary nonidentical
relation cannot be converted into identical relation. However,
from this nonidentical relation it can be obtained identical relation.

It appears that this is only possible under degenerate transformation.

\subsection*{Degenerate transformations. Obtaining identical relation
from nonidentical one}

To obtain the identical relation from the evolutionary nonidentical
relation, it is necessary that the closed exterior differential form
should be derived from the evolutionary differential form included into
evolutionary  relation.

However, as it has been shown above, the evolutionary form cannot be a closed form.
For this reason the transition from evolutionary form is possible only
to the {\it inexact} closed exterior form that is defined only on
pseudostructure.

To the pseudostructure it is assigned a closed dual form
(whose differential vanishes). For this reason the transition
from the evolutionary form to the closed inexact exterior form proceeds
only when the conditions of vanishing the dual form differential are
realized, in other words, when the metric form differential or
commutator becomes equal to zero.

The metric form of original manifold, on which the evolutionary
form (and correspondingly, nonidentical relation) is defined, is
not closed. The transition from  unclosed metric form with nonzero
differential to  closed metric form with the differential being
equal to zero is possible only as a degenerate transformation,
that is, as a transformation which does not conserve the
differential.

The conditions of vanishing the dual form differential (vanishing the
metric form commutator) are the conditions of degenerate
transformation.

Vanishing the metric form commutator under degenerate transformation points
to emergence of closed metric form, and this corresponds  to realization
of pseudostructure (the dual form).

Vanishing one term of the evolutionary form commutator, namely, the metric
form commutator, leads to that the second term of the commutator also
vanishes. This is due to the fact that the terms of the evolutionary form
commutator correlate between one another. The evolutionary form commutator
and, correspondingly, the differential, vanish on pseudostructure, and this
means that it appears the unclosed inexact exterior form.

If the conditions of degenerate transformation are realized, from
the unclosed evolutionary form, which differential is nonzero
$d\omega^p\ne 0 $,
one can obtain a differential form closed
on pseudostructure. The differential of this form equals zero. That is,
it is realized the transition

$d\omega^p\ne 0 \to $ (degenerate transformation) $\to d_\pi \omega^p=0$,
$d_\pi{}^*\omega^p=0$

On the pseudostructure $\pi$ evolutionary relation (1.10 ) transforms into
the relation
$$
d_\pi\psi=\omega_\pi^p\eqno(1.13 )
$$
which proves to be an identical relation. Indeed, since the form
$\omega_\pi^p$ is a closed one, on the pseudostructure this form turns
out to be a differential of some differential form. In other words,
this form can be written as $\omega_\pi^p=d_\pi\theta$. Relation (1.13 )
is now written as
$$
d_\pi\psi=d_\pi\theta
$$
There are differentials in the left-hand and right-hand sides of this relation. This means that the relation is an identical one.

Thus, it is evident that under degenerate transformation the
identical on pseudostructure relation can be obtained from the evolutionary
nonidentical relation. This is due to the realization of closed metric form, 
and correspondingly to the realization of closed exterior form, and this points 
to the emergency of pseudostructure and a conservative quantity on pseudostructure. 
The pseudostructure with conservative quantity makes up a differential-geometrical 
structure (the properties of such structure will be defined below).

Under degenerate transformation the evolutionary form differential
vanishes only {\it on pseudostructure}. This is an interior (being equal to zero)
differential. The total differential of the
evolutionary form is nonzero. The evolutionary form remains to be
unclosed, and for this reason the original
relation, which contains the evolutionary form, remains to be nonidentical.

Under realization of new additional conditions a new identical relation can
be obtained. As a result, the nonidentical evolutionary relation can generate
identical relations.

(It can be shown that all identical relations of the exterior differential
form theory are obtained from nonidentical relations by applying
the degenerate transformation.)

The conditions of degenerate transformation, that is, additional conditions, can be
realized, for example, if it will appear any symmetries of the metric form coefficients  or their derivatives.
{\it This can happen under selfvariation of the nonidentical relation.}

(While describing material system the conditions of degenerate
transformation are connected with degrees of freedom of a material
system).

To the conditions of degenerate transformation there
corresponds a requirement that some functional expressions become equal
to zero. Such functional expressions are Jacobians, determinants,
the Poisson brackets, residues, and others.

Mathematically the degenerate transformation is realized as a transition from one
frame of reference to another (nonequivalent) frame of reference. This is
a transition from the frame of reference connected with the manifold whose
metric forms are unclosed to the frame of reference being connected with
pseudostructure. The first frame of reference cannot be inertial or locally-
inertial frame. The evolutionary form and nonidentical evolutionary relation
are defined in the noninertial frame of reference. But the closed exterior form obtained
and the identical relation are obtained with respect to the
locally-inertial frame of reference.

\bigskip
{\bf Relation between degenerate transformations of exterior forms and
nondegenerate transformations of evolutionary forms.}
In the theory of closed exterior forms only nondegenerate transformations,
which conserve the differential, are used. Nondegenerate transformations of
evolutionary forms are transformations that do not conserve the differential.
Nevertheless, these transformations are mutually connected. The degenerate
transformations execute a transition from original deforming manifold
to pseudostructures. And the nondegenerate transformations execute a transition
from one pseudostructure to another. As the result of degenerate transformation
the closed inexact exterior form arises, and this points to origination of the
differential-geometrical structure. But under nondegenerate transformation
the transition from one closed form to another takes place, and this points
to the transition from one differential-geometrical structure to another.
When describing the evolutionary processes in material media using evolutionary
forms it will be shown that the degenerate transformation describes an
origination of physical structures (which are differential-geometrical structures)
in material media and that nondegenerate transformation describes a transition
from one physical structure to another.

\subsection*{Mechanism of realization of conjugated objects and operators.}

The mathematical apparatus of evolutionary differential forms
reveals the process of mutual topological variation of the
evolutionary form and the manifold. Under such variation the conditions
of degenerate transformation can be realized, and the exterior differential
form closed on the pseudostructure can arise (spontaneously).

To the closed exterior forms it can be assigned conjugated operators,
whereas to the evolutionary forms there correspond nonconjugated 
operators.
The transition from the evolutionary form to the closed exterior form
is that from nonconjugated operators to conjugated ones. This is expressed
as the  transition from the nonzero differential (unclosed evolutionary 
form) to the differential (closed exterior form) that equals zero.

Here it should be emphasized that the properties of deforming manifolds 
and skew-symmetric differential forms on such manifolds, namely, 
evolutionary forms, play a principal role in the process of conjugating.

The process of conjugating includes the following points:

1) selfvariation of nonidentical relation, namely, mutual variations of the
evolutionary form coefficients (which have
the algebraic nature) and of manifold characteristics (which have the
geometric nature) described by nonidentical evolutionary equation, and
2) realization of the degenerate transformation.

Hence one can see that the process of
conjugating is a mutual exchange between the quantities
of different nature and the degenerate transformation under additional
conditions. Here it should be pointed that the condition of degenerate
transformation (vanishing some functional expressions like, for example,
Jacobians, determinants and so on) may be realized spontaneously
while selfvarying the nonidentical relation if any symmetries appear.
It is possible if the system (that is described by this relation) possesses
any degrees of freedom.

One can see that the process of realization of conjugated operators or objects
is described by the nontraditional mathematical apparatus, namely, by
nonidentical relations and degenerate transformations.

As it has been shown above, closed exterior forms appear in many
mathematical formalisms. Practically all conjugated objects are
explicitly or implicitly connected with closed exterior forms. And
yet it can be shown that closed exterior forms are generated by
evolutionary differential forms, which are skew-symmetric
differential forms defined on the deforming varying manifolds.

\bigskip
By comparing the exterior and evolutionary forms one can see that they
possess the opposite properties. Yet it was shown that the
evolutionary differential forms generate the closed exterior
differential forms. This elucidates that the exterior and
evolutionary differential forms are unified whole.

\subsection{Evolutionary forms: Realization of differential-geometrical
structures and forming pseudometric and metric manifolds}

\subsection*{Realization of differential-geometrical structures.}

The process of realization of closed exterior form relates to the 
realization of closed metric form, which is a form being dual with
respect to the closed exterior form. The closed exterior form and
corresponding dual form make up a new conjugated object. Since the
closed exterior form is a conservative quantity (because the
differential of closed exterior form equals zero) and the dual form
describes the pseudostructure, such conjugated object (closed
exterior form and dual form) describes the
differential-geometrical structure, namely, the pseudostructure
with conservative quantity. As it has been already noted, such
structures are the example of the differential-geometrical
G-structures. (The physical structures, which forms physical
fields, and corresponding conservation laws, are just such structures).

The mathematical apparatus of evolutionary differential forms
describing the process of generation of closed inexact
exterior differential forms, describes thereby the process of
origination of the differential-geometrical structure.

Obtaining differential-geometrical structures is a process of
conjugating the objects. Such process is, firstly,
a mutual exchange between the quantities of different nature
(for example, between the algebraic and geometric quantities or between
the physical and spatial quantities), and, secondly, the establishment
of exact correspondence (conjugacy) of these objects. This process 
is described by selfvariation of nonidentical relation and 
degenerate transformation.

While describing the process of realization of the 
differential-geometrical structure the following questions arise. 
What does generate such structures, and what is the reason of such 
processes? These problems will be discussed in the next sections while 
studying physical processes in material media (which are 
controlled by conservation laws).

Below it is shown by what are defined the characteristics of the 
structures originated, how are such structures classified, how are 
formed the pseudometric and metric manifolds from these structures.

\bigskip
{\bf Characteristics of the differential-geometrical structures
realized.} For the sake of convenience in the subsequent
presentation the differential- geometrical structures, to which
inexact exterior and dual forms obtained from the evolutionary
forms correspond, will be called the binary structures or
Bi-Structures.

Since the closed exterior and dual differential forms, which
correspond to Bi-Structure arisen, were obtained from
the nonidentical relation that involves the evolutionary form, it is 
evident that the characteristics of such structure have to be connected: 

a) with those of the evolutionary form and of the deforming manifold
on which this form is defined, 

b) with the values of commutators of the
evolutionary form and the manifold metric form, and 

c) with the conditions of degenerate transformation as well.

The condition of degenerate transformation corresponds to a realization 
of the closed metric (dual) form and defines the pseudostructure.

Vanishing the interior commutator of the evolutionary form 
(on pseudostructure) corresponds to a realization of the closed (inexact) 
exterior form and points to emergence of conservative (invariant) quantity.

When Bi-Structure originates, the value of the total 
commutator of the evolutionary form containing two terms is nonzero. 
These terms define the following characteristics of Bi-Structures:

a) the first term of evolutionary form commutator (which is composed of 
the derivatives of the evolutionary form coefficients) defines the value of 
the discrete change of conservative quantity, that is, the quantum,
which the quantity conserved on the pseudostructure undergoes at the 
transition from one pseudostructure to another;

b) the second term (which is composed of the derivatives of
coefficients of the metric form connected with the manifold)
specifies the characteristics of Bi-Structures, which fixes the
character of the initial manifold deformation taking place before 
Bi-Structures  arose. (Spin is such an example). 
(This characteristics fixes the deformation of original manifold that
proceeded in the process of originating the differential-geometrical 
structure and was described by selfvariation of nonidentical relation).

Thus, the conditions of degenerate transformation determine
the pseudostructures;
the first term of the evolutionary form commutator determines
the value of the discrete change (the quantum) of conservative quantity; 
the second term of the evolutionary form commutator specifies the 
characteristics that fixes the character of the initial manifold 
deformation.

The discrete (quantum) change of conservative quantity proceeds in the 
direction that is normal to the pseudostructure. (Jumps of the 
derivatives normal to the potential surfaces are examples of such 
changes.)

Above it has been noted that the evolutionary form and the nonidentical
relation are obtained while describing the physical processes that 
proceed in material systems. For this reason it is evident that
the characteristics of Bi-Structure must also be connected with
the characteristics of the material system being described.

\bigskip
{\bf Classification of differential-geometrical structures realized.}
The closed forms that correspond to Bi-Structures are generated by the
evolutionary relation which includes the evolutionary form of $p$ degree.
Therefore, the structures originated can be classified by the parameter 
$p$.

The other parameter is the degree of closed forms generated by
the nonidentical evolutionary relation.

The nonidentical evolutionary relation with the forms of degree $p$ will
be called the evolutionary relation of degree $p$. Under degenerate 
transformation from {\it nonidentical} relation of degree $p$  
{\it an identical} relation is obtained. This identical relation 
contains the closed form of degree $k=p$.

To find how the parameter $k=p$ changes, one has to consider the problem of
integration of the nonidentical evolutionary relation.

The identical relation with closed form of degree $p$ obtained can be
integrated, since the right-hand side of such a relation can be
expressed in terms of differential (as well as the left-hand side).

By integrating the identical relation obtained one can
get the nonidentical relation of degree $(p-1)$.

The nonidentical relation of degree $(p-1)$ obtained can be integrated
once again if the corresponding degenerate transformation is realized, 
and the identical relation is formed. This identical relation will 
already include the closed form of degree $k=p-1$.

In this manner the evolutionary nonidentical relation of degree $p$ may
generate closed (on the pseudostructure) exterior forms of sequential
degrees $k=p, \dots, k=0$ and corresponding identical relations.

Thus, one can see that Bi-structures, to which there are assigned
the closed (on the pseudostructure) exterior forms, can depend on two
parameters. These parameters are the degree of evolutionary form $p$
(in the evolutionary relation) and the degree of created closed
forms $k$.

In addition to these parameters, another parameter appears,
namely, the dimension of space. If the evolutionary relation
generates the closed forms of degrees $k=p$, $k=p-1$, \dots,
$k=0$, to them there are assigned the pseudostructures of
dimensions $(N-k)$, where $N$ is the space dimension.

\subsection*{Forming pseudometric and metric manifolds.}

At this point it should be noted that at every
stage of the evolutionary process it is realized only one element of
pseudostructure, namely, a certain minipseudostructure.
(The example of minipseudoctructure is the wave
front. The wave front is an eikonal surface (the level surface), i.e.
a surface with conservative quantity.)

While varying the evolutionary variable the minipseudostructures form
the pseudostructure.

Manifolds with closed metric forms are formed by pseudostructures. They
are obtained from the deforming manifolds with unclosed metric forms. 
In this case the initial deforming manifold (on which the evolutionary 
form is defined) and the manifold with closed metric forms originated 
(on which the closed exterior form is defined) are different spatial 
objects.

It takes place the transition from the initial (deforming) manifold
with unclosed metric form to the pseudostructure, namely, to the
manifold with closed metric forms created. Mathematically this 
transition (the degenerate transformation) proceeds as
{\it a transition from one frame of reference to another, nonequivalent,
frame of reference.}

The pseudostructures, on which the closed {\it inexact} forms are
defined, form the pseudomanifolds.

To the transition from pseudomanifolds to metric space it is assigned
the transition from closed {\it inexact} differential forms 
to {\it exact} exterior differential forms.

It was shown above that the evolutionary relation of degree $p$ can
generate (with using the degenerate transformations) closed forms
of degrees $0,...,p$.  While generating closed forms of sequential
degrees  $k=p, k=p-1,..., k=0$ the pseudostructures of dimensions
$(n+1-k)$ are obtained. As a result of transition to the exact closed
form of zero degree the metric structure of the dimension $n+1$ is
obtained. 

Sections of the cotangent bundles (Yang-Mills fields), cohomologies 
by de Rham, singular cohomologies, pseudo-Riemannian and 
pseudo-Euclidean spaces, and others are examples of the psedustructures 
and spaces that are formed by pseudostructures. Euclidean and Riemannian 
spaces are examples of metric
manifolds that are obtained when changing to the exact forms.
Here it should be noted that the examples of pseudometric spaces are
potential surfaces (surfaces of a simple layer, a double layer and so
on). In these cases the type of potential surfaces is connected with
the above listed parameters.

Conservative quantities (closed exterior inexact forms) defined on
pseudomanifolds (closed dual forms) constitute some fields. 
(The physical fields are the examples of such fields.) The fields of 
conservative quantities are formed from closed exterior forms at the 
same time when the manifolds are created from the pseudoctructures.

Since the closed metric form is dual with respect to some closed exterior
differential form, the metric forms cannot become closed by themselves,
independently of the exterior differential form. This proves that
the manifolds with closed metric forms are connected with the closed
exterior differential forms.
This indicates that the fields of conservative quantities are
formed from closed exterior forms at the same time when the manifolds are created 
from the pseudoctructures. The specific feature of manifolds with closed metric 
forms that have been formed is that they can carry some information.
That is, the closed exterior differential forms and manifolds, on which 
they are defined, are mutually connected objects.

\bigskip
One can see that the evolutionary forms possess the properties, which 
enable one to describe the evolutionary processes, namely, the processes 
of generating the differential-geometrical structures and forming 
manifolds. In other mathematical formalisms there are no such 
possibilities that the mathematical apparatus of evolutionary and 
exterior skew-symmetrical forms possesses.

The evolutionary differential forms, which generate the closed
exterior forms, disclose the mechanism of origination of physical
structures forming physical fields  and the determinacy of these
processes.

Before describing a role of the evolutionary differential forms in
mathematical physics and field theory, it is necessary to show the
relation between the closed exterior forms and existing invariant
field theories.

\section{Role of exterior forms in field theory}

The role of exterior forms in field theory is due to the fact that they
reflect the properties of conservation laws.

\subsection{Conservation laws for physical fields (exact
conservation laws)}

Owing to the development of science the concept of ``conservation laws"
in thermodynamics, physics and mechanics contains different meanings.

In areas of physics related to the field theory and in the theoretical
mechanics ``the conservation laws" are those according to which there
exist conservative physical quantities or objects. These are the
conservation laws that above were called  ``exact".

In mechanics and physics of continuous media the concept of ``conservation laws"
is related to the conservation laws for energy, linear momentum, angular momentum,
and mass that establish the balance between the change of physical quantities
and external action. These are balance conservation laws.

In thermodynamics the conservation laws are associated with the principles
of thermodynamics.

Below it will be shown the relation between exact and balance
conservation laws and the connection of conservation laws with
the principles of thermodynamics.

\bigskip
{\it The exact conservation laws are those that state an existence of
conservative physical quantities or objects. The exact 
conservation laws are related to physical fields} \{The physical fields [9] 
are a special form of the substance, they are carriers of various interactions
such as electromagnetic, gravitational, wave, nuclear and other kinds of
interactions.\}

The closed exterior differential forms correspond to the exact conservation
laws. Indeed, from the closure conditions of the exterior differential 
form (see formulas (1.4), (1.6), (1.7)) it is evident that the closed 
exterior differential form is a conservative quantity. In this case
the closed inexact exterior differential form and the corresponding dual
form describe a conservative object, namely, there is a conservative
quantity only on some pseudostructure $\pi $. From this one can see that
the closed exterior differential form can correspond to the exact
conservation law.

The closure conditions for the exterior differential form 
($d_{\pi }\,\theta ^p\,=\,0$)
and the dual form ($d_{\pi }\,^*\theta ^p\,=\,0$) are
mathematical expressions of the exact conservation law.

It has been pointed above that the pseudostructure (dual form) and
the conservative quantity (the closed exterior form) define the
differential-geometrical (binary) structure (Bi-structure), which
is the example of G-Structure. It is evident that such structure
corresponds to the exact conservation law.

It is such structures (pseudostructures with a conservative physical
quantity) that correspond to exact conservation law, that are, the
physical structures from which physical fields are formed.

Equations for the physical structures ($d_{\pi }\,\theta ^p\,=\,0$,
$d_{\pi }\,^*\theta ^p\,=\,0$) turn out to coincide with the mathematical
expression for the exact conservation law.

The mathematical expression for the exact conservation law and its connection
with physical fields can be schematically written in the following manner:
$$
\def\\{\vphantom{d_\pi}}
\cases{d_\pi \theta^p=0\cr d_\pi {}^{*\mskip-2mu}\theta^p=0\cr}\quad
\mapsto\quad
\cases{\\\theta^p\cr \\{}^{*\mskip-2mu}\theta^p\cr}\quad\hbox{---}\quad
\hbox{physical structures}\quad\mapsto\quad\hbox{physical fields}
$$

It should be emphasized that the closed {\it inexact exterior forms}
correspond to the physical structures that form physical fields.
The {\it exact exterior forms} correspond to the material system
elements. (About this it will be said below).

It can be shown that the field theories, i.e. the theories
that describe physical fields, are based on the invariant and metric
properties of the closed exterior differential and dual forms
that correspond to exact conservation laws.

\subsection{Closed exterior differential forms in invariant field theories.
(Exact conservation laws and specific features of existing field theories)} 

The properties of closed exterior differential forms correspond to the
conservation laws for physical fields. Therefore, the mathematical 
principles of the theory of closed exterior differential forms lie at 
the basis of existing field theories.

The equations that are equations of the existing field theories are 
those obtained on the basis of the properties of the exterior 
differential form theory.

The Hamilton formalism is based on the properties of closed exterior and
dual forms of the first degree. The corresponding equation of the field has
the form:
$${{\partial s}\over {\partial t}}+H \left(t,\,q_j,\,p_j\right )\,=\,0,\quad
{{\partial s}\over {\partial q_j}}\,=\,p_j \eqno(2.1)$$ 
where $s$ is the field function (the state function) for the action functional
$S\,=\,\int\,L\,dt$. Here $L(t,\,q_j,\,\dot q_j)$ is the Lagrange
function, $H$ is the Hamilton function:
$H(t,\,q_j,\,p_j)\,=\,p_j\,\dot q_j-L$, $p_j\,=\,\partial
L/\partial \dot q_j$. To this equation it is assigned the closed
exterior form of the first degree, which is the Poincare invariant
$ds=-Hdt+p_j dq_j$. The closure conditions for this form and
corresponding dual form constitute the Hamiltonian systems:
$${{dq_j}\over {dt}}\,=\,{{\partial H}\over {\partial p_j}}, \quad
{{dp_j}\over {dt}}\,=\,-{{\partial H}\over {\partial q_j}}\eqno(2.2)$$

The Schr\H{o}dinger equation in quantum mechanics is an analog to
equation
(2.1) (where the conjugated coordinates are changed by operators),
and the Heisenberg equation is an analog to the appropriate integral of 
the equation (2.2) that have the form of canonical relations.
The conjugacy of Dirac's {\it bra-} and
{\it cket-} vectors in  quantum mechanics corresponds to the closure
condition of the zero degree exterior form [10]. \{The duality of closed forms
manifests
itself in the approaches by Schr\H{o}dinger and Heisenberg. Whereas
the zero degree closed exterior form corresponds to the Schr\H{o}dinger
equation, the closed dual form corresponds to the Heisenberg equation.
It can be pointed out that, whereas
the equations by Shr\H{o}dinger and Heisenberg describe the behavior of
the potential obtained from the zero degree closed form, Dirac's {\it bra-} 
and {\it cket}- vectors constitute the zero degree closed exterior form itself 
as the result of conjugacy (vanishing the scalar product)\}.

It is evident that the closed exterior and dual
forms of zero degree correspond to quantum mechanics.

The properties of closed exterior and dual forms of the second
degree lie at the basis of the electromagnetic field equations. The Maxwell
equations may be written as
$d\theta^2=0$, $d^*\theta^2=0$, where $\theta^2=
\frac{1}{2}F_{\mu\nu}dx^\mu dx^\nu$ (here $F_{\mu\nu}$ is the strength tensor).

Closed exterior and dual forms of the third degree correspond to the
gravitational field.

From the above said one can see that to each type of physical fields
there corresponds a closed exterior form of appropriate degree. (However,
to the physical field of given type it can be assigned closed forms of less
degree. In particular, to the Einstein equation for gravitational field it
corresponds the first degree closed form, although it was pointed out that
the type of a field with the third degree closed form
corresponds to the gravitational field.)

The connection between field theory and closed exterior differential
forms supports the invariance of field theory.

And here it should underline that field theories are based on the properties
of closed {\it inexact} forms. This is explained by the fact that only inexact
exterior forms  can correspond to the physical structures that form
physical fields.  The condition that the closed exterior forms,
which constitute the basis of field theory equations, are inexact ones
reveals in the fact that essentially all existing field theories include
a certain elements of noninvariance, i.e. they are based either on functionals
that are not identical invariants (such as Lagrangian, action functional, entropy)
or on equations (differential, integral, tensor, spinor, matrix and so on)
that have no identical invariance (integrability or covariance). Such elements
of noninvariance are, for example, the nonzero value of the curvature tensor in
Einstein's theory [3], the indeterminacy principle in Heisenberg's theory,
the torsion in the theory by Weyl [3], the Lorentz force in electromagnetic
theory [11], an absence of general integrability of the Schr\H{o}dinger
equations, 
the Lagrange function in the variational methods, an absence of the identical
integrability of the mathematical physics equations, and that of identical
covariance of the tensor equations,
and so on. Only if we assume elements of noncovariance, we can obtain
closed {\it inexact} forms that correspond to physical structures.

And yet, the existing field theories are invariant ones because they are
provided with additional conditions under which the invariance or covariance
requirements have to be satisfied. It is possible to show that these
conditions are the closure conditions of exterior or dual forms.
Examples of such conditions may be the identity relations: canonical
relations in the Schr\H{o}dinger equations, gauge invariance in electromagnetic
theory, commutator relations in the Heisenberg theory,
symmetric connectednesses,
identity relations by Bianchi in the Einstein theory, cotangent bundles in
the Yang-Mills theory, the Hamilton
function in the variational methods, the covariance conditions in the tensor
methods, etc. The field theory postulates are the expression of such conditions.

The connection between the field theory equations and closed
exterior forms shows that to every physical field it is assigned
the appropriate degree of closed exterior form. The type of gauge
transformations used in field theory depends on the degree of
closed exterior differential form.

This shows that it is possible to introduce a classification of physical
fields according to the degree of closed exterior form.
But within the framework of only exterior differential forms one cannot
understand how this classification is explained. (This can be elucidated
only by application of evolutionary differential forms.)
\bigskip

The exterior differential forms, whose properties correspond to the 
conservation laws, constitute the basis of the invariant field theories. 

The existing field theories allow to describe the physical fields.
However, because these theories are invariant ones
they cannot answer the question about the mechanism of originating
physical structures that form physical fields. The origination of physical
structures and forming physical fields are evolutionary processes,
and hence they cannot be described by the invariant field theories. Only
evolutionary theory can do this. The theory of exterior and 
evolutionary forms can be such a theory.

\section{Role of evolutionary forms in mathematical physics and field theory} 

The role of evolutionary forms in mathematical physics and field theory
(as well as the role of exterior forms) is due to the fact that they reflect
the conservation laws. However, these conservation laws are those not for physical 
fields but for material media. These are balance conservation laws.

\subsection{Balance conservation laws}

{\it The balance conservation laws are those that establish the balance between
the variation of a physical quantity and the corresponding external action.
These are the conservation laws for the material systems (material media)}.

The balance conservation laws are the conservation laws for energy, linear
momentum, angular momentum, and mass.

In the integral form the balance conservation laws express the following 
[12]: a change of a physical quantity in an elementary volume over a 
time interval is counterbalanced by the flux of a certain quantity 
through the boundary surface and by the action of sources. Under 
transition to the differential expression the fluxes are changed by 
divergences.

The equations of the balance conservation laws are differential (or integral)
equations that describe a variation
of functions corresponding to physical quantities [8, 12 - 14].
If the material system is not a dynamical one (as in the case
of a thermodynamic system), the equations of the balance conservation
laws can be written in terms of increments of physical quantities and
governing variables.

(The specific forms of these equations for thermodynamical and gas
dynamical material systems and the systems of charged particles
will be presented in the Appendix).

But it appears that, even without knowledge of the concrete form
of these equations, with the help of the differential forms one can see
specific features of these equations that elucidate the properties of
the balance conservation laws. To do so it is necessary to study the conjugacy
(consistency) of these equations.

\{The necessity of studying the conjugacy of the equations describing
any process has a physical meaning. If these equations
(or derivatives with respect to different variables) be not conjugated,
the solutions to corresponding equations prove to be noninvariant:
they are functionals rather then functions. The realization of the conditions
(while varying variables), under which the equations become conjugated ones,
leads to that the relevant solution becomes invariant. It will be shown below
that the transition to the invariant solution, which can be obtained
only using evolutionary forms, describes the mechanism of
evolutionary transition from one quality to another, which leads to emergence
of physical structures\}.

Equations are conjugate if they can be contracted into identical
relations for the differential, i.e. for a closed form.

Let us analyze the equations
that describe the balance conservation laws for energy and linear momentum.

We introduce two frames of reference: the first is an inertial one
(this frame of reference is not connected with the material system), and
the second is an accompanying
one (this system is connected with the manifold built by
the trajectories of the material system elements). The energy equation
in the inertial frame of reference can be reduced to the form:
$$
\frac{D\psi}{Dt}=A \eqno(3.1)
$$
where $D/Dt$ is the total derivative with respect to time, $\psi $ is the
functional
of the state that specifies the material system, $A$ is the quantity that
depends on specific features of the system and on external energy actions onto
the system. \{The action functional, entropy, wave function
can be regarded as examples of the functional $\psi $. Thus, the equation
for energy presented in terms of the action functional $S$ has a similar form:
$DS/Dt\,=\,L$, where $\psi \,=\,S$, $A\,=\,L$ is the Lagrange function.
In mechanics of continuous media the equation for
energy of an ideal gas can be presented in the form [13]: $Ds/Dt\,=\,0$, where
$s$ is entropy. In this case $\psi \,=\,s$, $A\,=\,0$. It is worth noting that
the examples presented show that the action functional and entropy play the
same role.\}

In the accompanying frame of reference the total derivative with respect to
time is transformed into the derivative along the trajectory. Equation 
(3.1)
is now written in the form
$$
{{\partial \psi }\over {\partial \xi ^1}}\,=\,A_1 \eqno(3.2)
$$
here $\xi^1$ is the coordinate along the trajectory.

In a similar manner, in the
accompanying frame of reference the equation for linear momentum appears
to be reduced to the equation of the form (see, for example, Appendix)
$$
{{\partial \psi}\over {\partial \xi^{\nu }}}\,=\,A_{\nu },\quad \nu \,=\,2,\,...\eqno(3.3)
$$
where $\xi ^{\nu }$ are the coordinates in the direction normal to the
trajectory, $A_{\nu }$ are the quantities that depend on the specific
features of the system and external (with respect to local domain)
force actions.

Eqs. (3.2), (3.3) can be convoluted into the relation
$$
d\psi\,=\,A_{\mu }\,d\xi ^{\mu },\quad (\mu\,=\,1,\,\nu )\eqno(3.4)
$$
where $d\psi $ is the differential
expression $d\psi\,=\,(\partial \psi /\partial \xi ^{\mu })d\xi ^{\mu }$.

Relation (3.4) can be written as
$$
d\psi \,=\,\omega \eqno(3.5)
$$
here $\omega \,=\,A_{\mu }\,d\xi ^{\mu }$ is the skew-symmetrical differential form of the first degree.

Since the balance conservation laws are evolutionary ones, the relation
obtained is also an evolutionary relation.

Relation (3.5) was obtained from the equation of the balance
conservation laws for
energy and linear momentum. In this relation the form $\omega $ is that of the
first degree. If the equations of the balance conservation laws for
angular momentum be added to the equations for energy and linear momentum,
this form in the evolutionary relation will be the form of the second degree.
And in  combination with the equation of the balance conservation law
of mass this form will be the form of degree 3.

Thus, in the general case the evolutionary relation can be written as
$$
d\psi \,=\,\omega^p \eqno(3.6)
$$
where the form degree  $p$ takes the values $p\,=\,0,1,2,3$..
(The evolutionary
relation for $p\,=\,0$ is similar to that in the differential forms, and it was
obtained from the interaction of energy and time.)

In relation (3.5) the form $\psi$ is the form of zero degree. And in relation
(3.6) the form $\psi$ is the form of $(p-1)$ degree.

Let us show that {\it the evolutionary relation  obtained from the equation
of the balance conservation laws proves to be nonidentical}.

To do so we shall analyze relation (3.5).

In the left-hand side of evolutionary relation (3.5) there is a
differential that is a closed form. This form is an invariant
object. The right-hand side of relation (3.5) involves the differential form
$\omega$, that is not an invariant object because in real processes, as it is
shown below, this form proves to be unclosed.

For the form to be closed the differential of the form or its commutator
must be equal to zero.

Let us consider the commutator of the
form $\omega \,=\,A_{\mu }d\xi ^{\mu }$.
The components of the commutator of such a form can be written as follows:
$$
K_{\alpha \beta }\,=\,\left ({{\partial A_{\beta }}\over {\partial \xi ^{\alpha }}}\,-\,
{{\partial A_{\alpha }}\over {\partial \xi ^{\beta }}}\right )\eqno(3.7)
$$
(here the term  connected with the manifold metric form
has not yet been taken into account).

The coefficients $A_{\mu }$ of the form $\omega $ have been obtained either
from the equation of the balance conservation law for energy or from that for
linear momentum. This means that in the first case the coefficients depend
on the energetic action and in the second case they depend on the force action.
In actual processes energetic and force actions have different nature and appear
to be inconsistent. The commutator of the form $\omega $ constructed from
the derivatives of such coefficients is nonzero.
This means that the differential of the form $\omega $
is nonzero as well. Thus, the form $\omega$ proves to be unclosed and is not
a measurable quantity.

This means that the relation (3.5)
involves an unmeasurable term. Such a relation cannot be an identical
one.

Hence,  without the knowledge of particular expression for the form
$\omega$,
one can argue that for actual processes the relation obtained from the equations
corresponding to the balance conservation laws proves to be
nonidentical.

In similar manner it can be shown that  general relation (3.6) is also
nonidentical.

As it was noted the differential forms $\omega^p$ and relations (3.6)
are defined on accompanying manifold, that is, on the manifold made up by
the trajectories of elements of material system. This manifold is deforming one.
The metric forms of such manifold cannot be closed. The differential forms
defined on such manifold are evolutionary differential forms, which have been
described in the fourth section.

The differential forms $\omega^p$ are evolutionary forms, and relations
(3.5) and (3.6) are examples of nonidentical relations for evolutionary
forms.

In the Appendix the derivation of nonidentical relations for thermodynamic
and gas dynamic systems as well as for the system of charged particles is
presented and their brief analysis is given.

Thus, one can see that to the conservation laws for physical fields
(exact conservation laws) there correspond the properties of closed inexact
forms, whereas to the conservation laws for material media there correspond
the properties of evolutionary differential forms.

The nonidentity of the relation obtained from the equations of balance
conservation laws means that the equations of balance conservation laws turn
out to be nonconjugated (thus, if from the energy equation we
obtain the  derivative of $\psi $ in the direction along the trajectory
and from the momentum equation we find the derivative of $\psi $
in the direction normal to the trajectory and then we calculate
their mixed derivatives, from the condition that
the commutator of the form $\omega $ is nonzero it follows that
the mixed derivatives prove to be noncommutative).

The nonconjugacy of the equations of balance conservation laws
reflects the properties of balance conservation laws,
namely, their noncommutativity. This property have a governing
importance for the evolutionary processes.

\subsection{Evolutionary process in material medium and origination of physical
structures}

In this subsection the mechanism of evolutionary
processes in material media, which are accompanied by the emergence of
physical structure, is described on the basis of the mathematical
apparatus of evolutionary differential forms.

The conservation laws are shown to play a governing role
in evolutionary processes.

[To emphasize a connection between the
mathematical and physical principles, some of the principles listed above
will be included into the titles of some subsections. It will be used
double titles,
namely, those having the physical meaning and those having the
mathematical one. The corresponding mathematical principles are
presented in brackets.]

\subsection*{Nonequilibrium of the material system. (Nonidentity
of the evolutionary relation)}

It was shown above that the evolutionary relation that was obtained from 
the balance conservation law equations proves to be nonidentical.
This points to the noncommutativity of the balance conservation laws. By
analyzing the behavior of the nonidentical evolutionary relation one can
understand to what result the noncommutativity of the balance
conservation laws leads.

Before we start analyzing the evolutionary relation some relevant
concepts such as a ``local domain" of a material system,
``accompanying manifolds", ``nonequilibrium" and ``locally equilibrium"
states of a material system are to be explained.

{\it The local domain of material system.}
The local domain of material system is the element and its vicinity.
In deriving the evolutionary relation
with the first degree form (for the balance conservation laws of energy
and linear momentum) it was considered the local domain of material
system that involves the material system element and its vicinity, this
is, the material system element was an element of the local domain.
For the evolutionary relation
with the second degree form the local domain for the evolutionary
relation with the first degree form will serve as the element of the
local domain, and so on. For the evolutionary relation with zero degree
form the element of the material system will serve as the local domain
(rather than the element).

{\it Accompanying manifold.}
The accompanying manifold is the manifold constructed by
the trajectories of the elements of corresponding local domains. For the
evolutionary relation with $p=1$ the accompanying manifold is the
manifold formed from the trajectories of the material system elements
because in this case the elements
of the material system in itself serve as the elements of local domain.
It is to be noted that the accompanying manifold
is constructed of trajectories of the elements of the {\it local domain},
rather than of {\it material system} (see the concept of a ``local
domain").

{\it Nonequilibrium and locally equilibrium states of material system.} 
It is evident that the nonequilibrium is connected with actions of some
forces. One has to distinguish between  external forces and internal ones (in
the subsequent analysis one will also deal with potential forces).
Here  external forces (external actions) mean the forces that act
onto the local domain. Internal forces are those that act inside the
local domain of the material system.
The nonequilibrium is connected with action of {\it internal forces}.
The nonequilibrium state is the state of a material system when internal forces act
in each local domain. If there are no internal forces, the state of material
system is in equilibrium. If there are no internal forces only in a particular
domain of the material system but there are internal forces in the neighboring
local domains, such a state of the material system  will be referred
to as the state ``in local equilibrium". In this case the total state of the
material system is not equilibrium.
\bigskip

We are coming now to the analysis of the evolutionary relation.

It was mentioned above that the noncommutativity of the balance
conservation laws is connected with a state of the material system. This
is reflected by the evolutionary relation.

It should be emphasized that the evolutionary relation treats a state
of the {\it local domain} of the material system.

\{The evolutionary relation takes into account the interaction of an element of
the local domain with its vicinity.
The noncommutativity of the balance conservation laws points to the fact that the
element and its vicinity prove to be nonconjugated.\}

Let us consider  evolutionary relation (3.5).

If the evolutionary relation proves to be identical, one can
obtain the differential $d\psi $ and find the state function $\psi $, 
this will indicate that the material system state is in equilibrium. But if the
evolutionary relation be nonidentical, this indicates an absence of the
differential $d\psi $ and nonequilibrium of the material system state.
(Hereafter the differential $d\psi $ will be called the state differential as
it specifies the material system state. This is a closed form. If the state
differential be an exact closed form, this corresponds to the equilibrium
system state, whereas, if the state differential be an inexact closed form,
this will correspond to the locally equilibrium state).

The evolutionary relation gives a possibility to determine
either presence or absence of the differential (the closed form).
And this allows us, firstly, to recognize whether the material system
state is in equilibrium, in local equilibrium or not in equilibrium,
and secondly, to determine the conditions of transition from one state
into another (this explains the mechanism of such a transition).
If it is possible to determine the differential $d\psi$ from
the evolutionary relation, this indicates that the system is in equilibrium
or locally equilibrium state. And if the differential cannot be determined,
then this means that the system is in a nonequilibrium state.

It is evident that if the balance conservation laws be commutative,
the evolutionary relation would be identical and from that it would be possible
to get the differential $d\psi $, this would indicate  that the material system
is in the equilibrium state.

However, as it has been shown, in real processes the balance conservation laws
are noncommutative. The evolutionary relation is not identical and from this
relation one cannot get the differential $d\psi $. This means that the system
state is nonequilibrium.

The nonequilibrium state means that there is an internal force in the
material system. It is evident that the internal force originates at the
expense of some quantity described by the evolutionary form commutator.
(If the evolutionary form commutator be zero, the evolutionary relation
would be identical, and this would point to the equilibrium state, i.e.
the absence of internal forces.) Everything that gives a contribution
into the evolutionary form commutator leads to emergence of the internal
force. 

Thus, the nonidentity of the evolutionary relation obtained from the equations 
of the balance conservation laws points to the noncommutativity of the balance
conservation laws and the nonequilibrium material system state produced
as a result. A quantity described by the evolutionary differential form
commutator serves as the internal force.

Each external action, as the result of which a change of physical
quantities has been produced, has the nature different from
that of the material system itself. For this reason the changed physical
quantities cannot directly become the physical quantities of the material
system itself. (The noncommutativity of the balance conservation laws does
not allow a direct transition of the external actions into the physical
quantities of the material system). The changed physical quantities prove to
be inconsistent. As a result it arises an unmeasurable quantity that
is described by the commutator of the evolutionary form $\omega^p$
and acts as an internal force.

To become the consistent physical quantities of the material system itself,
the modified physical quantities have to come to agreement with the
properties of the material system. Such transitions, as it will be shown below,
are also governed by the balance conservation laws.

\subsection*{Selfvariation of nonequilibrium state of material system.
(Selfvariation of the evolutionary relation)}

What does the material system nonequilibrium indicated
by the nonidentity of the evolutionary relation results in?

While describing the properties of evolutionary forms it has been shown 
that, the evolutionary and nonidentical relation is a selfvarying one.

Such a specific feature of the evolutionary relation explains the 
particulars of the material system, namely, {\it the selfvariation of 
its nonequilibrium state}.

The selfvariation mechanism of the {\it nonequilibrium}
state of the material system can be understood if analyze the selfvariation
of the evolutionary relation go on. For this purpose we have to analyze
the topological properties of the evolutionary form commutator.

The evolutionary form in the evolutionary relation is defined on the
accompanying manifold that for real processes appears to be the deformable
manifold because it is formed simultaneously with a change of the
material system state and depends on the physical processes.
Such a manifold cannot be a manifold with closed metric forms.
Hence, the term containing the characteristics of the
manifold will be included into the evolutionary form commutator in
addition to the term connected with derivatives of the form
coefficients. The interaction between these terms of different nature
describes a mutual change of the state of the material system.

Let us examine this with an example of the commutator of the form
$\omega \,=\,A_{\mu }\,d\xi ^{\mu }$ that is included into
evolutionary relation (3.5).

We assume that at the beginning the accompanying manifold
was that with the first degree closed metric form. In this case
the commutator of the form $\omega$ can
be written as (3.7). If at the next instant any action affects
the material system, this commutator turns out to be nonzero. The state
of the material system becomes nonequilibrium and it will arise an internal
force whose action will lead to a deformation of the accompanying manifold.
The metric form commutator of the accompanying manifold, which specifies the
deformation, will become nonzero (that is, the metric form of the accompanying
manifold will be unclosed).  In the commutator of the form $\omega $ it will
appear an additional term, that specifies a deformation
of the manifold and is a commutator of the manifold metric form. {If it is
possible to define the coefficients of connectedness
$\Gamma_{\alpha \beta }^{\sigma }$ (for a nondifferentiable
manifold they are skew-symmetric ones), the form commutator
may be written as (1.3), where $A_{\alpha} = a_{\alpha}$.

The emergence of the second term can only change the commutator and
cannot make it zero (because the terms of the commutator have
different nature). The further deformation (torsion) of the manifold
will go on. This leads to a change of the metric form commutator,
produces a change of the evolutionary form and its commutator and so on.
Such a process is governed by the nonidentical evolutionary relation and,
in turn, produces a change of the evolutionary relation.

The process of selfvariation of the evolutionary relation points
to a change of the material system state. But the material system state
remains nonequilibrium in this process because the internal forces
do not vanish due to the evolutionary form commutator remaining nonzero.

At this point it should be emphasized that such selfvariation of the material
system state proceeds under the action of internal (rather than external)
forces. That will go on even in the absence of external forces. That is,
the selfvariation of the
nonequilibrium state of the material system takes place.

Here it should be noted that in a real physical process the internal forces
can be increased (due to the selfvariation of the nonequilibrium state of the
material system).  This can lead to the development of instability in the
material system [15].
\{For example, this was pointed out in the works by Prigogine
[16]. ``The excess entropy" in his
works is analogous to the commutator of a nonintegrable form for the
thermodynamic system.  ``Production of excess entropy" leads to the
development of instability\}.

\subsection*{Transition of the material system into a locally equilibrium
state. Origination of the physical structures.
(Degenerate transformation. Emergence of closed exterior forms)}

Thus, it was shown that in real processes the material
system is in a nonequilibrium state (with an internal force).
This follows from the analysis of the nonidentical evolutionary relation 
obtained from the balance conservation law equations.

Now the question arises whether the material system can got rid of
the internal force and transfer into the equilibrium state?

The internal force is described by the evolutionary form commutator.
But the evolutionary form commutator cannot vanish.
This means that the internal force, which is described by the evolutionary
form commutator, cannot disappear. That is, the material system cannot be
transformed into the equilibrium (without internal forces) state.

However, the material system can change from the nonequilibrium state
into the locally equilibrium state. This follows from the evolutionary
differential form properties. Under degenerate transformation the 
identical relation can be obtained from the nonidentical evolutionary 
relation. That is, from nonidentical relation (3.6) it is obtained 
the identical on pseudostructure relation
$$
d_\pi\psi=\omega_\pi^p\eqno(3.8)
$$
where the form $\omega_\pi^p$ is one closed on pseudostructure.

The identical relation obtained from the nonidentical evolutionary
relation under degenerate transformation integrates the state
differential and the closed inexact exterior form. The
availability of the state differential $d_\pi\psi$ indicates that
the material system state becomes a locally equilibrium state
(that is, the local domain of the system under consideration
changes into the equilibrium state). The availability of the
exterior closed inexact form $\omega_\pi^p$ means that the
physical structure is present. This shows that the transition of
material system into the locally equilibrium state is accompanied
by the origination of physical structures.

The conditions of degenerate transformation are connected with
symmetries that can be obtained from the coefficients of
commutators of evolutionary and metric forms. Such symmetries can
be due to the degrees of freedom of material system and its
elements. The translational degrees of freedom, internal degrees
of freedom of the system elements, and so on can be examples of
such degrees of freedom.

As it was noted above, to the degenerate transformation it must
correspond  vanishing of some functional
expressions, such as Jacobians, determinants, the Poisson
brackets, residues and others. Vanishing these
functional expressions is the closure condition for a dual form.
And it should be emphasize once more that {\it the degenerate 
transformation is realized as the transition from the accompanying 
noninertial coordinate system to the locally inertial system}.

The availability of the degrees of freedom in the material system indicates 
that it is allowed the degenerate transformation, which, in turns, allows 
the state of the material systems to be transformed from a nonequilibrium 
state to a locally equilibrium state.
But, for this to take place in reality it is necessary that the
additional conditions
connected with the degrees of freedom of the material system be
realized. It is selfvariation of the nonequilibrium state of the material
system describe by the selfvarying evolutionary relation that could give 
rise to realization of the additional conditions. This can appear only
spontaneously because it is caused by internal (rather than external)
reasons (the degrees of freedom are the characteristics of
the system rather than of external actions).

Under degenerate transformation the evolutionary form differential vanishes
only along a certain direction. In other words, the interior
differential equal to zero is realized. But in this case the total differential
of the evolutionary form is nonzero. The evolutionary form commutator
does not vanish. Vanishing the interior differential of the evolutionary form means
that there is a closed inexact form, and this points to the locally equilibrium
state of the material system.
At the same time a nonzero value of the total differential
of evolutionary form means that the form remains unclosed. This shows
that the total state of the material system remains nonequilibrium.

Thus, from the properties of the nonidentical evolutionary relation and
those of the evolutionary form one can see that under realization of the
additional condition (which is a condition of degenerate transformation)
the transition of the material system state from nonequilibrium
to locally equilibrium state can be realized. Such transition is accompanied
by emergency of physical structures.
It should be emphasized once again that such a transition can occur only
spontaneously.

Here we should recall once again that the closed inexact form is a quantity
with double meaning, namely, it is both the conservative quantity
and the measurable quantity that acts as a potential force.
The transition of the material system from nonequilibrium into
a locally equilibrium state (which is indicated by the formation of
a closed form) means that the unmeasurable quantity described by the nonzero
commutator of the nonintegrable differential form $\omega^p$, that acts as an
internal force, transforms into the measurable quantity. It is evident that
it is just the measurable quantity that acts as a potential force. In other
words, the internal force transforms into a potential force.

\bigskip
Thus, the mathematical apparatus of the evolutionary differential forms
elucidates a mechanism of the evolutionary process in material media
and of the emergency of physical structures. This mechanism involves
the following steps.

1) The external actions onto the material system are transformed
into the unmeasurable quantity that acts as an internal force and brings the material system into
the nonequilibrium state. ({\it The nonzero value
of the evolutionary form commutator. Nonidentity of the evolutionary relation
obtained from the balance conservation laws}).

2) Selfvariation of the nonequilibrium state of the material system.
The deformation of accompanying manifold.
({\it Selfvariation of the nonidentical evolutionary relation. The topological
properties of the evolutionary form commutator}).

3) Realization of the degrees of freedom of the material system
in the process of selfvariation
of the nonequilibrium state of the system itself. ({\it
Degenerate transformations}).

4) Transition of the material system from the nonequilibrium state into
the locally equilibrium one: the transition of an internal force
into a potential force. The emergence of physical
structures. ({\it Formation of closed inexact forms and obtaining
the differential $d_\pi\psi$ that specifies the state of the material system}).

\{Here it should be emphasized the following.

The evolutionary relation nonidentity that follows from the
conservation law noncommutativity just reflects the
overdeterminacy of the set of the balance conservation law
equations.  Actually, a number of the balance conservation law
equations is equal to a number of desired physical quantities that
specify the material system. But since the physical
quantities relate to the same material system, it has to be
some connection between them. (This connection is executed by the
function that specifies the system state.) And as the physical
quantities are related to each other, then the set of the balance
conservation law equations proves to be overdetermined one. A
realization of the additional conditions, when from the
nonidentical evolutionary relation it follows the identical
relation, there corresponds to that from the overdetermined set of
equations it results the set of consistent equations from which
one can find the desired physical quantities. As the additional
conditions may be realized only in the discrete manner, then the
solutions to this set may be only quantized.

In the book by A.Pais ``The Science and the Life of Albert Einstein"
the author wrote: ``He (Einstein) hoped that the idea of the
overdeterminacy will lead to
getting the discrete solutions. He also believed that from the future theory
it will be possible to derive the partly localized solutions that would
correspond to particles that carries the quantized electric charge".\}

\section{Evolutionary forms: Properties of  physical structures.
Formation of physical fields and manifolds}

In Section 3 the  mechanism of evolutionary processes in material media
was described, and it was shown that the evolutionary processes lead
to origination of the physical structures. (These are just such structures that
form physical fields.)

In the present Section the properties of the physical structures and their
connection with material media are described. The mechanism of forming physical fields
and corresponding manifolds is described.

\subsection{Characteristics of physical structures.
(Characteristics of differential forms)}

The exterior form closed on the pseudostructure in combination
with the dual form determining the pseudostructure constitute the
differential-geometrical structure named the Bi-Structure.

The physical structure is such differential-geometrical structure.

The physical structure is an object obtained by conjugating the
conservative physical quantity, which is described by inexact
closed exterior form, and the pseudostructure, which is described
by relevant dual form.

What characteristics do the physical structures possess?

The closed exterior forms corresponding to physical structures are
conservative quantities. These conservative quantities describe
certain charges.

Under transition from one structure to another
the conservative quantity corresponding to the closed exterior form
discretely changes, and the pseudostructure also changes discretely.

Discrete changes of the conservative quantity and pseudostructure are
determined by the value of the evolutionary form commutator, which the
commutator has at the instant when the physical structure originates.
The first term of the evolutionary form commutator obtained from
the derivatives of the evolutionary
form coefficients controls the discrete change of the conservative
quantity. The second one obtained from the derivatives of the metric
form coefficients of the initial manifold controls
the pseudostructure change.

Spin is the example of the second characteristic. Spin is a characteristic 
that determines a character of the manifold deformation before 
origination of the quantum. (The spin value depends on the form degree.)

Breaks of the derivatives of the potential along the direction normal
to the potential surface, breaks of the derivative in transition
throughout the characteristic surfaces and in transition throughout the
wave front, and others are the examples of discrete change of the
conservative quantity.

A discrete change of the conservative quantity and that of the
pseudostructure produce the quantum that is obtained while going from
one structure to another.
The evolutionary form commutator formed at the instant of the structure
origination determine characteristics of this quantum. 

\subsection{Connection between  physical structures
originated and material systems. (Identical relation: connection 
between the state differential and the closed inexact exterior form)}

Since closed inexact exterior forms corresponding to physical structure
are obtained from the evolutionary relation for the material system, it 
follows that physical structures are generated by the material systems. 
(This is controlled by the conservation laws.)
The closed exterior forms obtained correspond to the
state differential for material system. The differentials of entropy,
action, potential and others are the examples of such differentials.

In this manner the physical structures are connected with the material
system, its elements, its local domains. The characteristics of physical
structure are determined by the characteristics of the material
system that generates these physical structures.

The equation of the pseudostructure (dual form) is obtained, as it has 
been shown, from the condition of degenerate transformation.
And this relates to the degrees of freedom of the material system.

The characteristics of inexact closed exterior forms are defined
by the characteristics of evolutionary forms following from the balance
conservation laws for material media and, hence, depend on the 
characteristics of material media.

Here it should call attention to one more fact. The material
system can generate physical structures only if the system is in
nonequilibrium state, that is, if it experience the influence of
any actions. As it has been shown, the internal forces arisen are
described by the evolutionary form commutator. Therefore, the
characteristics of the physical structures arisen will also depend
on the characteristics of the evolutionary form commutator
describing any actions onto the material system.

\bigskip
In material system the origination of physical structure reveals 
as a new measurable and observable formation that
spontaneously arises in material system.
$\{${\it As the examples it can be fluctuations, pulsations,
waves, vortices, and creating massless particles.}$\}$.

In the physical
process this formation is spontaneously extracted from the local
domain of material system and so it allows the local domain of material
system to get rid of an internal force and come into the locally equilibrium state.

The formation created in a local domain of material system
(at the cost of unmeasurable quantity that acts in the local domain
as an internal force)
and liberated from that, begins acting onto the neighboring local domain
as a force. This is a potential force, this fact is indicated by the double
meaning of the closed exterior form (on the one hand, a
conservative quantity, and, on other hand, a potential force).
(This action was produced by the material system in itself,
and therefore this is a potential action rather than an arbitrary one).

The neighboring domain of the material system works over this action
that appears to be external with respect to that. If in the process the
conditions of conjugacy of the balance conservation laws turn out to be
satisfied again, the neighboring domain
will create a formation by its own, and this formation will be extracted
from this domain. (If the conjugacy conditions are not realized, the
process is finished.) In such a way the formation can move
relative to the material system. (Waves are the example of such motions).

The extraction of a formation from the local domain of the system
is accompanied by emergence of the break surfaces in the material system.
The contact breaks are the examples of such surfaces. These breaks do
not propagate relative to the material system as shocks or shock waves
do. The black holes may be such break surfaces.

The observed formation and the physical structure are not identical objects.
If the wave be such a formation, then the wave front is the physical structure.
In this case the wave element is a minipseudostructure.

How is the created formation  connected with a change of physical quantities of
the material system?

Assume that at some instant the local domain of the material system was in
equilibrium. That is,
its physical quantities, for example, energy and momentum were
consistent and simultaneously measurable physical quantities. Then under
the effect of external (with respect to the local domain) actions the physical
quantities were changed and ceased to be consistent measurable quantities. 
When the degrees of freedom of the
material system are realized, this allows the physical quantities, which
were changed at the expense of external actions, to redistribute in
such a way as to become measurable physical quantities, namely,
the inherent (corresponding to the nature of material system) quantities of
material system.

It is evident that the transition from the initial measurable physical
quantities to new physical quantities realized is discrete one.

That is, the proper measurable physical quantities are changed discretely.
This discrete change of physical quantities is revealed as a formation 
created.

\subsection*{Potential forces. (Duality of closed exterior forms as
conservative quantities and as potential forces)}

As it was shown above, the unmeasurable quantity, that acts as an 
internal force and has been stored at the cost of all external 
actions giving contribution into the commutator, is converted into 
a measurable quantity that acts as a potential force. This is indicated 
by the availability of a closed inexact form that can correspond to 
the potential force.

Where, from what, and on what does the potential force act?

The potential force is an action of the created (quantum) formation onto
the local domains of the material system over which it is translated.
And if the internal force acts in the interior of the local domain of 
the material system (and it caused it to deform), the potential force 
acts onto the neighboring domain. The local domain gets rid of its 
internal force and modifies it into a potential force that acts onto the neighboring
domains. An unmeasurable quantity, that acts in a local domain as an internal
force, is transformed into a measurable quantity of the observable 
formation (and the physical structure as well) that is emitted from 
the local domain and acts onto the neighboring domain as a force equal 
to this quantity.

The potential forces, as well as the internal forces,
originate at the cost of
the external actions, but the potential forces (unlike interior
ones) are connected
with the measurable quantities.

Unlike arbitrary external forces, the potential forces are those that
originate at the expense of external actions processed by the material 
system.

If the external actions equal zero (the evolutionary form commutator be 
equal to zero), then internal and potential forces equal zero.

Thus, one has to recognize the forces of three types: 1) external forces
(external actions),
2) internal forces that originate in local domains of material system
due to the fact that the physical quantities of the material system 
changed by external actions turn out to be inconsistent, and 3) the 
potential forces are forces of the action of the formations 
(corresponding to physical structures) onto a material system.

The potential forces are the source of originating internal forces in 
the neighboring domains of the material system, on which the potential 
forces act (such as the forces that are external with respect to that 
domain). For this reason the total state of a material system can 
remain nonequilibrium just
without additional external actions (nonpotential forces).

The potential force, whose value is conditioned by the
quantity of the commutator
of the evolutionary form $\omega^p$ at the instant of the formation
production, acts normally to the  pseudostructure, i.e. with
respect to
the integrating direction, along which the interior differential
(the closed form) is formed.
The potential forces are described, for example,
by jumps of the derivatives in the direction normal to the
characteristics, to the potential surfaces and so on (as well as the
physical structure characteristics).
This corresponds to the fact that the evolutionary form
commutators along these directions are nonzero.

The duality  of the closed inexact form as a conservative
quantity and as a potential force shows that
the potential forces are the action of formations corresponding to
the physical structures onto the material system.

Here the following should be pointed out. The physical structures are
generated by local domains of the material system. They are the elementary
physical structures. By combining with one another they can form the
large-scale structures and physical fields.

\subsection*{Characteristics of the formation created: intensity, vorticity, 
absolute and relative speeds of propagation of the formation. (Value
of the evolutionary form commutator, the properties of material system)}

As it was already mentioned, in the material system a created physical structure
is revealed as an observable formation. It is evident that the characteristics
of the formation, as well as those of the created physical structure, are
determined by the evolutionary form and its commutator and by the material system
characteristics.

Analysis of the formation and its characteristics allows
a better understanding of the specific features of physical structures.

Since  the formation is a result of converting an unmeasurable quantity 
described by the evolutionary form commutator
into a measurable physical quantity, it is evident that the intensity of
the formation created (as well as the discrete change
of the physical structure characteristics) is controlled by the quantity
that was stored by the evolutionary form commutator at the instant when the
formation appeared.

The first term of the commutator constructed of the
derivatives of the form coefficients controls the intensity of the formation,
whereas the second term that specifies the deformation of the
accompanying manifold (bending, torsion, curvature) is fixed as any internal
characteristics of the formation originated (which corresponds to 
vorticity, for example).

In the preceding subsection it was shown that the formation emerged
in the local domain of material system acts onto the neighboring local
domain of material system. Such action is determined by potential force.
It is evident that the formation intensity is revealed as a potential force.

The other characteristics of the formation obserrved are absolute and relative
speeds of the formation propagation.

As it was pointed out above, the observed formation that appears in 
material system moves along material system. The physical structure is
a front of such moving formation.

Here it should be emphasized once again that the moving formation at each
instant appears as a newly appeared formation.

The absolute propagation speed of the formation originated (speed in the
inertial frame of reference) is obtained from the condition of degenerate
transformation that corresponds to conjugacy of the balance conservation laws.
As it was already pointed out, these
conditions are determined by the material system characteristics
and are connected with the degrees of freedom of material system.

These conditions are the differential equations of pseudostructure. They
specify the rate of formatting pseudostructure, namely, the speed of the front 
of the formation emerged. This is just an absolute speed of the formation emerged. 

Here it should be emphasized that the speed of the formation originated, though
it is determined by the material system characteristics, is not a parameter
of the system itself. This is a quantity that at every instant while the formation moves
with respect to  material system is realized anew as the
condition of conjugacy of the balance conservation laws.

If the material system is homogeneous, the speed of translation will have
the same values. However, it is not constant because it is formed
anew at every instant of the evolutionary processes.

The relative speed (speed in the accompanying frame of reference) is
a speed of the formation translation  {\it relative to material system}.

The relative speed is equal to the absolute speed minus the velocity
of the local domain elements of material system (or of the elements of 
material system if $p=1$). That is, the relative speed of the observed 
formation is determined by the degrees of freedom of material system and by 
the velocities of the elements of local domain.

\bigskip
In such a way the following correspondence between the characteristics of the
formations emerged and characteristics of the evolutionary forms, of the
evolutionary form commutators and of the material system is established:

1) an intensity of the formation (a potential force)
$\leftrightarrow$ {\it the  value of the first term in the
commutator of nonintegrable form} at the instant when the formation
is created;

2) vorticity (an analog of spin)  $\leftrightarrow$ {\it the second term in the commutator
that is connected with the metric form commutator};

3) an absolute speed of propagation of the created formation (the
speed in the inertial frame of reference) $\leftrightarrow$ {\it additional
conditions connected with degrees of freedom of material system};

4) a speed of the formation propagation relative to material system
$\leftrightarrow $  {\it additional
conditions connected with degrees of freedom of material system
and the velocity of elements of local domain}.

%\bigskip
Analysis of formations originated in material system that correspond to
physical structures enables us to clarify some properties of physical
structures.

The rate of varying the pseudostructures that are described by dual forms
defines the absolute speed of formations.
That is, the pseudostructures are connected with the front of the
formation translation. Since a certain physical quantity is conserved on
the pseudostructure (the closed form), the pseudostructure is a level surface.
The equation of pseudostructure is the equation of eikonal surface.
(The eikonal is an example of physical structure.)

It can be shown that the equations of the characteristic surfaces, the surfaces
of potential (of simple layer, double layer), the residue equations and so on,
obtained from the equations of mathematical physics
serve as the equations for pseudostructures. \{In the papers [8,17] the
connection of equations for one, two, \dots\ eikonals with
the equations for characteristics, with the Hamilton equation, and others
was shown\}. 

The mechanism of creating the pseudostructures lies at the basis of
forming the pseudometric surfaces and their transition into metric spaces
(see the next subsection). It should be pointed out that the eigenvalues and 
the coupling constants appear as the conjugacy conditions for
exterior or dual forms, the numerical constants are the conjugacy
conditions for exact forms.

\subsection{Formation of pseudometric and metric spaces. (Integration of the
nonidentical evolutionary relation)}

The mechanism of forming pseudometric and metric spaces is connected with
the creation of pseudostructures.

In Subsection (1.5) it has been shown that the evolutionary nonidentical relation
containing the evolutionary form of degree $p$ may 
generate the pseudostructure and the closed (on the pseudostructure)
exterior forms of sequential degrees $k=p, \dots, k=0$. And yet the
pseudostructure dimensionality depends on the dimension of space
on which the exterior forms are defined.

What is implied by the concept ``space"?

While deriving the evolutionary relation two frames of reference were
used and, correspondingly, two spatial objects. The first frame of reference
is an inertial one, which is connected with the space where material
system is situated and is not directly connected with material system.
This is an inertial space, it is a metric space. (As it will be shown below,
this space is also formed by the material system itself.) The second frame of
reference is a proper one, it is connected with the accompanying manifold,
which is not a metric manifold.

As it is known, the form degree cannot be greater than the space dimension.
Therefore, if the dimension of the inertial space is $n$, the maximal degree
of the form will be $n$. For material system in such a space there work
the balance conservation laws that are convoluted into the evolutionary relation
of the degree $p$. In the real processes practically all conservation laws 
allowed in the space of a given dimension interact with one another. For 
given $n$ the value of $p$ will be practically a maximal value, that is, $p=n$.

Assume that $n=2$ and $p=2$. If the form $\omega^2$ is
an unclosed form, the commutator of this form will act in the
space of $n+i=2+i$ dimension (here the additional dimension is
denoted by an imaginary unit as it does not commutate
with other dimensions). This means that the deformed accompanying manifold
will not be imbedded into the original inertial space of dimension $n$.
This leads to that the dimension of the space formed increases by one.

It was shown above that the evolutionary relation of degree $p$
can generate (in the presence  of the degenerate transformations)
closed forms of the degree $0\le k\le p$ on the pseudostructures.
While generating closed forms of sequential degrees $k=p$,
$k=p-1$, \dots, $k=0$ the pseudostructures of dimensions
$(n+1-k)$: 1, \dots, $n+1$ are obtained. As a result of transition
to the exact closed form of zero degree the metric structure of
the dimension $n+1$ is obtained. Under influence of the 
external action (and in the presence of degrees of freedom) the
material system can transfer the initial inertial space into the
space of the dimension $n+1$. $\{$It is known that the
skew-symmetric tensors of the rank $k$ correspond to closed
exterior differential forms, and the pseudotensors of the rank
$(N-k)$, where $N$ is the space dimension, correspond to the
relevant dual forms. The pseudostructures correspond to such
tensors, but on the space formed with the dimension $n+1$. That
is, $N=n+1$\}.

Under the effect of external actions (and in the presence of degrees 
of freedom) the material system can convert the initial inertial space 
of the dimension $n$ into the space of the dimension $n+1$. Thus, 
a certain stage of forming the metric space is completed. Every material 
system has the cycle that includes four stages $(n=0,1,2,3)$. 
The cycle ends and a new cycle can begin. (This corresponds to one 
system being embedded into another one). The mechanism of formation of 
the pseudostructures and the metric structures can explain, in 
particular, how the internal structure of the elements of material 
system is formed. (See Subsection 6.1). 

So it can be seen that the inertial spaces are not absolute spaces where 
actions are developed, they are spaces generated by material systems.

The pseudo-Riemann and
pseudo-Euclidean spaces and others can be regarded as examples of 
pseudostructures and spaces that are formed in a similar manner. The
Riemann and Euclidean spaces are the example of metric manifolds
obtained in changing to exact forms. Below we present a brief analysis
of the space corresponding to gravitational field.

\subsection*{Space of gravitational field}

Material system (medium), which generates
gravitational field, is a cosmological system.
What can be said about the pseudo-Riemann manifold and Riemann space?
The distinctive property of the Riemann manifold is an availability of
the curvature. This means that the metric form commutator of the third
degree is nonzero. Hence, it does not equal zero the evolutionary form
commutator of the third degree $p=3$, which involves into itself the 
metric form commutator. That is, the evolutionary form that enters into 
the evolutionary relation is unclosed, and the relation is nonidentical.

When realizing pseudostructures of the dimensions $1$, $2$, $3$, and $4$ and
obtaining the closed inexact forms of the degrees $k=3$, $k=2$, $k=1$, and 
$k=0$, the pseudo-Riemann space is formed, and the transition to the 
exact form of zero degree corresponds to the transition to Riemann space.

It is well known that while obtaining the Einstein equations it
was suggested that there are fulfilled the conditions [3,18]: the
Bianchi identity is satisfied, the coefficients of connectedness
are symmetric, the condition that the coefficients of
connectedness are the Christoffel symbols, and an existence of the
transformation, under which the coefficients of connectedness
vanish. These conditions are the conditions of realization of the
degenerate transformations for nonidentical relations obtained
from the evolutionary relation of the degree $p=3$ and after changing 
to the exact relations. In this case to the Einstein equation
there corresponds the identical relations of the first degree.

The above described mechanism of forming the manifolds elucidates the connection
between the space and material objects. In his paper [19] S.Weinberg gives
more than one historical concepts of the space.
He wrote that the connection of the space with  material objects
was pointed out by Leibnitz who believed that there is no philosophical
necessity in any concept of space apart from that following from the
connections with material objects. In addition, S.Weinberg cites another
similar concept, namely, the Mach
principle,  which claims that in the definition of inertial system
the masses of Earth and celestial bodies play a role. The idea of physical
space as a continuum whose properties are governed by the matter was realized 
by A.Einstein.
The above described mechanism of formatting manifolds is one more substantiation 
of the connection of the space with material objects.

\subsection{Forming physical fields. Classification of physical structures.
(Parameters of closed and dual forms)}

Since the physical structures are generated by numerous local domains of 
material system and at numerous instants of realizing various degrees
of freedom of material system, it is evident that they can generate fields.
In this manner physical fields are formed.
To obtain the physical structures that form a given physical field, 
one has to examine the material system  corresponding to this field and 
the appropriate evolutionary relation. In particular, in the Appendix it is 
shown that, for to obtain the thermodynamical structures (fluctuations, 
phase transitions, etc), one has to analyze the
evolutionary relation for thermodynamical systems,
to obtain the gas dynamic ones (waves, jumps, vortices, pulsations)
one has to employ the evolutionary relation for gas dynamic
systems, for the electromagnetic field one must employ
a relation obtained from equations for charged particles.

Closed forms that correspond to physical structures are generated by
the evolutionary relation having the parameter $p$ that defines a number of
interacting balance conservation laws. Therefore, the physical structures
can be classified by the parameter $p$. The other parameter is a degree
of  closed forms generated by the evolutionary relation.
As it was shown above, the evolutionary relation of
degree $p$ can generate closed forms of degree $0\leq k \leq p$. Therefore,
physical structures can be classified by the parameter $k$ as well.
Closed exterior forms of the same degree realized in spaces of
different dimensions prove to be distinguishable because the dimension of the
pseudostructures, on which the closed forms are defined, depends on the space
dimension. As a result, the space dimension also specifies the physical
structures. This parameter determines the properties of physical structures
rather than their type.

Hence, from the analysis of the evolutionary relation one can see that the 
type and the properties of the differential-geometrical structures
and, consequently, of the physical structures (and, accordingly, of physical
fields) for a given material system depend on a number of interacting balance
conservation laws $p$, on the degree of closed forms realized $k$, and on the
space dimension. By introducing a classification with respect to $p$, $k$,
and space dimension we can understand an internal connection of various
physical fields and interactions. Such a connection will be considered in
Subsection (6.1).

\section{Conservation laws. Symmetries. Causality}

\subsection{Conservation laws}

Here it should be emphasized once again the role of the conservation laws
in evolutionary processes and the connection between the conservation 
laws for material systems and those for physical fields 
(connection between balance conservation
laws and exact conservation laws). 

From the description of evolutionary process it was seen that in the evolutionary
processes proceeded in material medium the balance conservation laws for energy,
linear momentum, angular momentum, and mass play a controlling and governing
role. Such a role of the conservation laws for material systems is due to its
peculiarities, that is, they turn out to be noncommutative ones.

This noncommutativity of balance conservation laws reflects the fact that
the external actions cannot directly be converted into quantities (measurable)
of material system, and hence, they produce a certain nonmeasurable quantity,
which acts as internal force and converts the material system into the
nonequilibrium state.

The interaction of the noncommutative balance
conservation laws controls the process of selfvarying
the nonequilibrium state of material system. Such a process can cause the
realization of the degrees of freedom of material system (if they
are available). The degrees of freedom allow the nonmeasurable quantity,
which acts as external force, to be converted into the measurable quantities
of the material system itself. The physical structures are the result
of realizing the measurable quantities of material system.
They appear in material system as the production of new formations. And yet
the material system is changed into the locally equilibrium state.

Since the exact conservation laws correspond to physical
structures, the origination of physical structure points to a realization
of the exact conservation law. From here one can see the
connection between the balance and exact conservation laws.

The physical structures that correspond to the exact
conservation laws are produced by material system in the evolutionary
processes, which are  based on the interaction of the noncommutative
balance conservation laws.

{\it Noncommutativity of the balance conservation laws
and their controlling role in the evolutionary processes
accompanied by emerging  physical structures practically
have not been taken into account in the explicit form anywhere}. The
mathematical apparatus of evolutionary differential forms enables one
to take into account and to describe these points.

\subsection{Symmetries}

The exterior and evolutionary skew-symmetric differential forms,
which describe the conservation laws, disclose thereby the properties
and specific features of symmetries.

The gauge symmetries, which are interior symmetries of the field
theory equations, are connected with the conservation laws for physical
fields. These  symmetries are those of (inexact) closed forms. 

The closure property of the exterior form means that any objects,
namely, elements of the exterior form, components of elements,
elements of the form differential, exterior and dual forms, and
others, turn out to be conjugated. And the conjugacy is possible
only if there are symmetries of one or other type.

The gauge transformations of field theory, which are nondegenerate
transformations of the closed exterior differential forms, are connected with
the gauge symmetries. Since the closed exterior differential form is a differential
(a total one if the form is exact, or an interior one on pseudostructure 
if the form is inexact), it is
obvious that the closed form proves to be invariant under all
transformations that conserve the differential. The unitary transformations
(0-form), the tangent and canonical transformations (1-form), the gradient and
gauge transformations (2-form) and so on are  examples of such transformations.
{\it These are gauge transformations for spinor, scalar, vector, and tensor
(3-form) fields}.

The internal symmetries in field theory are those of closed exterior
differential forms, whereas the external symmetries are symmetries of relevant
dual forms.

It has been shown that the closed exterior forms and relevant dual forms, which
correspond to the conservation laws for physical fields, are obtained from
the evolutionary forms, which describe the balance conservation laws
for material media. This proceeds under degenerate transformation, which is
connected with the degrees of freedom of material system. The conditions of
degenerate transformation defines the closed dual form. From this it follows
that the external symmetries, namely, the symmetries of dual forms, are due to
the degrees of freedom of material system. It is for this reason the exterior
symmetries are spatial-temporal symmetries.

The realization of the closed dual form,
which proceeds due to the degrees of freedom of material system,
leads to realization of the closed exterior form, that is, to the
conjugacy  of the differential form elements, and emergency of
internal symmetries. From this one can see the connection between
internal and external symmetries.

Whereas the internal symmetries are connected with the conservation laws for
physical fields, the external symmetries caused by the degrees of freedom of
material media are connected with the balance conservation laws for material
media.

The nondegenerate transformations are connected with internal symmetries, 
and the degenerate transformations of evolutionary forms
are connected with external symmetries.

\subsection{Causality}

The Encyclopedia gives the following definition of the concept of
``Causality". ``The causality is a genetic connection between separate states
of types and forms of matter in the process of its motion and development.
The emergency of any objects and systems and their time-developments have 
their own grounds in the preceding states of matter; these grounds are known 
as causes and the changes produced by them are referred to as consequences. 
The essence of the causality is a production of the consequence by the 
cause; the consequence, which is governed by the cause, produces the reverse action 
to the cause".

A determinacy of the above described evolutionary processes in material systems
matches with this definition of the causality.

It can be understood that all external actions onto material system lead to
the evolutionary processes. This is a cause of the evolutionary
processes. The emergency of physical structures and formation of physical fields
are the consequences of the evolutionary processes in material systems.

From the description of evolutionary processes one can see a connection between
physical fields and material systems. Physical fields are generated by
material systems. And the noncommutative balance conservation laws for material
systems control these processes.

The causality of evolutionary processes in material systems that lead
to emergency of physical structures is justified by the mathematical
apparatus of evolutionary and exterior differential forms describing the
balance conservation laws in material systems.

The evolutionary process in material system and the emergency of physical
structures can take place, if

1) the material system is subject to an external action ({\it the
evolutionary form commutator in the evolutionary relation obtained from the
balance conservation laws is nonzero}),

2) the material system possesses the degrees of freedom ({\it there
are the conditions of degenerate transformation, under which from the
nonidentical evolutionary relation an identical relation is obtained}),

3) the degrees of freedom of material system have to be realized,
that is possible only under selfvariations of the nonequilibrium state of 
material system ({\it the conditions of degenerate transformation
have to be satisfied, this is possible under selfvariation of the
nonidentical evolutionary relation}).

If these conditions (causes) are fulfilled, in the material system
physical structures arise ({\it this is indicated by the presence of closed
inexact form obtained from the identical relation}).

It is these structures that form physical fields.

Note that to each physical field it is assigned its own material system.
The question of what physical system does correspond
to a particular physical field is still an open question. Examples of such
material systems are the thermodynamic, gas dynamical, cosmological systems,
the system of charged particles, and so on. Maybe, for elementary particles
the physical vacuum is such a system.

The evolutionary process may lead to creation of elements of its own
material system
({\it while obtaining the exact form of zero degree}).

The emergency of physical structures in the evolutionary process proceeds
spontaneously and is manifested as an emergency of certain observable
formations. In this manner the causality of emerging various
observable formations in material media is explained. Such formations and their
manifestations are fluctuations, turbulent pulsations, waves, vortices,
creating massless particles, and others.

\bigskip
The existing field theories that are invariant ones are based on some
postulates. The investigation performed enables us to make the following
conclusion. The postulates, which lie at the basis of the existing field
theories, correspond to the conditions of conjugacy of the
balance conservation laws for material systems that generate physical
structures forming physical fields.

In physics there exists ``the causality principle" [9]. This principle
establishes the permitted limits of mutual influence of physical events. 
``The causality principle" is a statement that is significantly narrower than 
the general philosophic understanding of causality formulated above. It 
does not clarify the causal and consequential connection of physical 
phenomena.

\section{A certain aspects of quantum field theory and approaches to
general field theory} 

The importance of evolutionary forms in mathematical physics
and field theory consists in the fact that they disclose the
mechanism of forming physical fields and substantiates the
determinacy of corresponding processes. From the description of
evolutionary processes in material media using the evolutionary
skew-symmetric differential forms it follows that material media
(material systems) generate physical fields. And the conservation
laws, which turn out to be noncommutative, control these
processes.

Connection of physical fields with material media and the classification
of physical structures based on parameters that specify the conservation laws
for material media enable one to see internal relations of different physical
fields and their common foundations.

\subsection{On interactions and classification of physical structures
and physical fields}

As it was shown above, the type of physical structures (and,
accordingly, of physical fields) generated by the evolutionary
relation depends on the degree of differential forms $p$ and $k$
and on the dimension of original inertial space $n$ (here $p$ is
the degree of the evolutionary form in the nonidentical relation
that is connected with a number of interacting balance
conservation laws, and $k$ is the  degree of closed form
generated by the nonidentical relation). By introducing the
classification by numbers $p$, $k$, $n$ one can understand the
internal connection between various physical fields. Since physical 
fields are the carriers of interactions, such
classification enables one to see the connection between
interactions.

On the basis of the properties of evolutionary forms that correspond to
the conservation laws one can suppose that such a classification may be
presented in the form of the table given below.
This table corresponds to elementary particles.

\{It should be emphasized the following. Here the concept of ``interaction"
is used in a twofold meaning: an interaction of the balance conservation laws
that relates to material systems, and the physical concept of ``interaction"
that relates to physical fields and reflects the interactions of physical
structures, namely, it is connected with the exact conservation laws\}.

Recall that the interaction of balance conservation laws for energy and
linear momentum corresponds to the value $p=1$, with the balance
conservation law for angular momentum in addition this corresponds to
the value $p=2$, and with the balance conservation law for mass in addition
it corresponds to the value $p=3$. The value $p=0$ corresponds to interaction
between time and the balance conservation law for energy.

\bigskip
\centerline{TABLE}

%{\scriptsize
\noindent
\begin{tabular}{@{~}c@{~}c@{~}c@{~}c@{~}c@{~}c@{~}}
\bf interaction&$k\backslash p,n$&\bf 0&\bf 1&\bf 2&\bf 3

\\
\hline
\hline
\bf gravitation&\bf 3&&&&
	\begin{tabular}{c}
	\bf graviton\\
	$\Uparrow$\\
	electron\\
	proton\\
	neutron\\
	photon
	\end{tabular}

\\
\hline
	\begin{tabular}{l}
	\bf electro-\\
	\bf magnetic
	\end{tabular}
&\bf 2&&&
	\begin{tabular}{c}
        \bf photon2\\
	$\Uparrow$\\
	electron\\
	proton\\
	neutrino
	\end{tabular}
&\bf photon3

\\
\hline
\bf weak&\bf 1&&
	\begin{tabular}{c}
	\bf neutrino1\\
	$\Uparrow$\\
	electron\\
	quanta
	\end{tabular}
&\bf neutrino2&\bf neutrino3

\\
\hline
\bf strong&\bf 0&
	\begin{tabular}{c}
	\bf quanta0\\
	$\Uparrow$\\
	quarks?
	\end{tabular}
&
	\begin{tabular}{c}
	\bf quanta1\\
	\\

	\end{tabular}
&
\bf quanta2&\bf quanta3

	\\
\hline
\hline
	\begin{tabular}{c}
	\bf particles\\
	material\\
	nucleons?
	\end{tabular}
&
	\begin{tabular}{c}
	exact\\
	forms
	\end{tabular}
&\bf electron&\bf proton&\bf neutron&\bf deuteron?
\\
\hline
N&&1&2&3&4\\
&&time&time+&time+&time+\\
&&&1 coord.&2 coord.&3 coord.\\
\end{tabular}
%}

In the Table names of the particles created are given. Numbers placed near
particle names correspond to the space dimension. In braces \{\} the
sources of interactions are presented. In the next to the last row we
present massive particles (the elements of material system) formed by 
interactions (the exact forms of zero degree obtained by
sequential integrating the evolutionary relations with the evolutionary forms
of degree $p$ correspond to these particles). In the bottom row the dimension
of the {\it metric} structure created is presented.

From the Table one can see the correspondence between the degree $k$ of the
closed forms realized and the type of interactions. Thus, $k=0$ corresponds to
the strong interaction, $k=1$ corresponds to the weak interaction,
$k=2$ corresponds to the electromagnetic interaction, and $k=3$ corresponds
to the gravitational interaction.
The degree $k$ of the closed forms realized and the number of interacting
balance conservation laws determine the type of interactions and the type
of particles created. The properties of particles are governed by the space
dimension. The last property is connected with the fact that
closed forms of equal degrees $k$, but obtained from the evolutionary
relations acting in spaces of different dimensions $n$, are distinctive
because they are defined on the pseudostructures of different dimensions
(the dimension of the pseudostructure $(n+1-k)$ depends on the dimension
of initial space $n$). For this reason the realized physical structures
with closed forms of equal degrees $k$ are distinctive in their properties.

In the Table one cycle of forming physical structures is
presented. This cycle involves four levels, to each of which there
correspond their own values of $p$ ($p=0,1,2,3$) and space
dimension $n$. The structures formed in the previous cycle serve
as a source of interactions for the first level of new cycle. The
consequent cycles reflect the properties of material systems
subsequently imbedded. And yet a given level has specific
properties that are inherent characteristics of the same level in
another cycles. This can be seen, for example, from comparison of
the cycle described and the cycle in which to the exact forms
there correspond conductors, semiconductors, dielectrics, and
neutral elements. The properties of the elements of the third
level, namely, of neutrons in one cycle and of dielectrics in
another are identical to the properties of so called "magnetic
monopole" [20,21].

The Table presented provides the idea about the dimension of pseudostructures
and  metric structures.

In the bottom row of the Table the dimension $N$ of the metric structure
formed is presented.  From original space of the dimension $0$ the metric space
of the dimension $1$ (it can occurs to be time) can be realized. From space of 
the dimension $1$ the metric space of the dimension $2$ (time and
coordinate) can appear and so on. From original space of the dimension $3$
it can be formed the metric space of the dimension $4$ (time and $3$
coordinates). Such space is convoluted  and a new dimension cannot already be
realized. This corresponds to ending the cycle. (Such metric space with
corresponding physical quantity defined by the exact exterior form
is the element of new material system.)

The method of studying evolutionary processes developed on the basis of
exterior and evolutionary differential forms with using the properties of the
conservation laws may serve as an approach to general field theory.

\subsection{Mathematical apparatus of skew-symmetric differential
forms as the basis of general field theory}

At Section 2 it was shown the role of closed exterior
forms in field theory. The properties of closed exterior differential forms 
explicitly or implicitly manifest themselves essentially in all formalisms of
field theory, such as the Hamilton formalism, tensor approaches,
group methods, quantum mechanics equations, the Yang-Mills theory and
others.  

Such a role of closed exterior forms in field theory is related to the fact that these forms reflect the properties of the conservation laws for physical
fields. The differential-geometrical structures described by closed (inexact)
exterior forms and relevant dual forms are physical structures that make up
physical fields. The degrees of closed exterior forms set the classification
of physical fields and interactions. To the strong, weak, electromagnetic, and
gravitational interactions there correspond the closed exterior forms of zeroth,
first, second, and third degrees. Gauge transformations for spinor,
scalar, vector, and tensor fields are transformations of closed exterior forms
of zeroth, first, second, and third degrees.

Such properties of closed exterior forms can be useful in establishing 
the unified field theory. 

And the theory of skew-symmetric differential forms, which unites
the theory of closed exterior forms and the theory of evolutionary
forms (generating the closed inexact exterior forms), can serve as
an approach to the general field theory. Such a theory enables one
not only to describe physical fields, but also shows how are the
physical fields produced, what does generate them, what is a cause
of these processes.

The field theories that are based on exact conservation laws allow to
describe physical fields. However, because these theories are invariant ones
they cannot answer the question about the mechanism of originating
physical structures that form physical fields. The originating physical
structures and forming physical fields are evolutionary processes,
and hence they cannot be described by the invariant field theories. Only
evolutionary theory can do this. The mathematical apparatus of evolutionary
differential forms can serve as the mathematical apparatus of such theory.

The basic mathematical foundations that describe the evolutionary
process in material systems, and the mechanism of originating 
physical structures evidently must be included into the evolutionary
field theory.

It should be noted that it is rather difficult  to realize all
these mathematical foundations and in many cases this turns out to
be impossible. The difficulties may be caused by a derivation of
the evolutionary relation that describes the mechanism of
originating physical structures and forming physical fields. To do
this one has to know the equations of the balance conservation
laws for material systems that generate given physical fields.
The problems can be also caused by the fact that these equations
have to be written in the frame of reference that is related to
the deforming accompanying manifold. Moreover, this can lead to
difficulties in the process of obtaining closed exterior forms
from evolutionary relation, because for to do this it is necessary
to obtain the additional conditions that correspond to the degrees
of freedom of material system and have to be realized by
themselves when changing the evolutionary relation.

These problems may appear to be practically unsolvable.

However, a knowledge of the basic mathematical principles of
evolutionary theory may be helpful while studying the mechanism
of originating physical fields.

The results of qualitative investigations of evolutionary processes
using the mathematical apparatus of evolutionary differential
forms enables one to see the common properties that unify all physical
fields. The physical fields are generated by material media,
and at the basis of this it lies the interaction of the noncommutative
conservation laws of energy, linear momentum, angular momentum, and mass
for material media. This explains the causality of physical phenomena
and clarifies the essence of postulates that lie at the basis of
existing field theories.

These results enable one to understand what do the properties of physical
structures depend on and with which parameters is the classification of
physical structures connected, and hence to see the internal connections between various
physical fields.

The properties of physical structures depend primarily
on which material systems (media) generate physical structures
(but the physical structures generated by different material media
possess common properties as well).

One of the parameters, according to which it is possible to
classify physical structures and physical fields, is the number of
interacting balance conservation laws. This is the parameter $p$
that ranges from 0 to 3. (Recall that the case $p=1$ corresponds
to interaction of the balance conservation laws of energy and
linear momentum, the case  $p=2$  does to that of energy, linear
momentum, and angular momenta, the case  $p=3$  corresponds to
interaction of the balance conservation laws of energy, linear and
angular momenta, and mass, and to $p=0$ it corresponds an
interaction between time and the balance conservation law of
energy.) This parameter, which is the evolutionary form degree
that enters into the evolutionary relation, specifies a type of
physical fields. (So, the electromagnetic field is obtained from
interaction between the balance conservation laws of energy and
linear and angular momenta. The gravitational field is obtained as
the result of interactions between the balance conservation laws
of energy, linear momentum, angular momentum, and mass.)

The other parameter is the degree of closed differential forms realized 
from given evolutionary relation. The values of these parameters denoted  
by $\kappa$ range from
$p$ to $0$. This parameter, which corresponds to physical structures realized,
characterizes the connection between physical structures and 
exact conservation laws. It specifies a type of interaction of physical
fields.

Since the realization of physical structures proceeds discretely,
this emphasizes a quantum character of physical fields.

One more parameter is the dimension $n$ of the space in which the physical
structures are generated. This parameter points to the fact that the
physical structures, which belong to common type of the exact
conservation laws, can be distinguished by their space structure.
(The classification with
respect to these parameters not only elucidates the connections between the
physical fields generated by material media, but explains the mechanism
of creating the elements of material media themselves and demonstrates 
the connections between material media as well).

By comparison of the invariant and evolutionary approaches to field theory
one can state the following.
Physical fields are described by invariant field theory that is based on exact
conservation laws. The properties of closed exterior differential forms lie at
the basis of mathematical apparatus of the invariant theory. The
mechanism of {\it forming } physical fields can be described only by
evolutionary theory. The evolutionary theory that is based on
the balance conservation laws for material systems is just such a theory.
It is evident that as the general 
field theory it must serve a theory that involves the basic mathematical
foundations of the evolutionary and invariant field theories.
As an approach to such general field theory it can serve the theory of
skew-symmetric differential forms, which unites the theories of closed exterior forms
and evolutionary forms.

{\it The papers on the theory of evolutionary skew-symmetric differential
forms written in 2002-2005 years are in ArXive (http://arXiv.org/find)}

\bigskip
\rightline{\large\bf Appendix}

\bigskip
\centerline {\large\bf The analysis of balance conservation laws for } 
\centerline {\large\bf thermodynamic and gas dynamic systems and for} 
\centerline {\large\bf the system of charged particles} 

\subsection*{Thermodynamic systems}

The thermodynamics is based on the first and second principles of thermodynamics
that were introduced as postulates [7].
The first principle of thermodynamics, which can be written in the form
$$dE\,+\,dw\,=\,\delta Q\eqno(A.1)$$
follows from the balance conservation laws for energy and linear momentum
(but not only from the conservation law for energy). This is analogous to 
the evolutionary relation for the thermodynamic system. Since $\delta Q$ 
is not a differential, relation (A.1) which corresponds to the first principle 
of thermodynamics, as well as the evolutionary relation, appears to be a 
nonidentical relation.
This points to a noncommutativity of the balance conservation
laws (for energy and linear momentum) and to a nonequilibrium state of the
thermodynamic system.

If condition of the integrability be satisfied, from the nonidentical
evolutionary relation, which corresponds to the first principle of
thermodynamics, it follows an identical relation. It is an identical relation
that corresponds to the second principle of thermodynamics.

If $dw\,=\,p\,dV$,  there is the integrating factor
$\theta$ (a quantity which depends only on the characteristics of the system),
where $1/\theta\,=\,pV/R$ is called the temperature $T$ [7].
In this case the form $(dE\,+\,p\,dV)/T$ turns out to be a differential
(interior) of some quantity that referred to as entropy $S$:
$$(dE\,+\,p\,dV)/T\,=\,dS \eqno(A.2)$$

If the integrating factor $\theta=1/T$ has been
realized, that is, relation (A.2) proves to be satisfied, from relation 
(A.1),
which corresponds to the first principle of thermodynamics,
it follows
$$dS\,=\,\delta Q/T \eqno(A.3)$$
This is just the second principle of thermodynamics for reversible processes.
This takes place when the heat input is the only action onto the system.

If in addition to the heat input the system experiences a certain mechanical
action
(for example, an influence of boundaries), we obtain
$$dS\, >\,\delta Q/T \eqno (A.4)$$
that corresponds to the second principle of thermodynamics for irreversible
processes.

In the case examined above the differential of entropy (rather than entropy
itself) becomes a closed form. $\{$In this case entropy  manifests itself
as the thermodynamic potential, namely, the function of state. To the
pseudostructure there corresponds
the state equation that determines the temperature dependence on the
thermodynamic variables$\}$.

\subsection*{Gas dynamical systems}

We take the simplest gas dynamical system, namely, a flow of ideal
(inviscid, heat nonconductive) gas [13].

Assume that the gas (the element of gas dynamic system) is a thermodynamic 
system in the state of local equilibrium (whenever the gas dynamic system 
itself may be
in nonequilibrium state), that is, it is satisfied the relation [7]
$$Tds\,=\,de\,+\,pdV \eqno(A.5)$$
where $T$, $p$ and $V$ are the temperature, the pressure and the gas
volume, $s$ and $e$ are entropy and internal energy per unit volume.

Let us introduce two frames of reference: an inertial one that is not connected
with material system and an accompanying frame of reference that is connected
with the manifold formed by the trajectories of the material system elements.

The equation of the balance conservation law of energy for ideal gas can
be written as [13]
$${{Dh}\over {Dt}}- {1\over {\rho }}{{Dp}\over {Dt}}\,=\,0 \eqno(A.6)$$
where $D/Dt$ is the total derivative with respect to time (if to denote 
the spatial coordinates by $x_i$ and the velocity components by $u_i$,
$D/Dt\,=\,\partial /\partial t+u_i\partial /\partial x_i$). Here  $\rho=1/V $
and $h$ are respectively the mass and the entalpy densities of the gas.

Expressing entalpy in terms of internal energy $e$ using the formula
$h\,=\,e\,+\,p/\rho $ and using relation (A.5), the balance conservation law
equation (A.6) can be put to the form
$${{Ds}\over {Dt}}\,=\,0 \eqno(A.7)$$

And respectively, the equation of the balance conservation law for linear
momentum can be presented as [13,22]
$$\hbox {grad} \,s\,=\,(\hbox {grad} \,h_0\,+\,{\bf U}\times \hbox {rot} {\bf U}\,-{\bf F}\,+\,
\partial {\bf U}/\partial t)/T \eqno(A.8)$$
where ${\bf U}$ is the velocity of the gas particle,
$h_0=({\bf U \cdot U})/2+h$, ${\bf F}$ is the mass force. The operator $grad$
in this equation is defined only in the plane normal to the trajectory.

Since the total derivative with respect to time is that along the trajectory,
in the accompanying frame of reference equations (A.7) and (A.8)
take the form:
$${{\partial s}\over {\partial \xi ^1}}\,=\,0 \eqno (A.9)$$
$${{\partial s}\over {\partial \xi ^{\nu}}}\,=\,A_{\nu },\quad \nu=2, ... \eqno(A.10)$$
where $\xi ^1$ is the coordinate along the trajectory,
$\partial s/\partial \xi ^{\nu }$
is the left-hand side of equation (A.8), and $A_{\nu }$ is obtained from the
right-hand side of relation (A.8).

Equations (A.9) and (A.10) can be convoluted into the equation
$$ds\,=\,A_{\mu} d\xi ^{\mu}\eqno(A.11)$$
where $\,A_{\mu} d\xi ^{\mu}=\omega\,$ is the first degree differential form
(here $A_1=0$,$\mu =1,\,\nu $).

Relation (A.11) is the evolutionary relation for gas dynamic system
(in the case of local thermodynamic equilibrium). Here $\psi\,=\,s$.
$\{$It worth notice that in the evolutionary relation for thermodynamic
system the dependence of entropy on thermodynamic variables is investigated
(see relation (A.5)), whereas in the evolutionary relation for gas dynamic
system the entropy dependence on the space-time variables is considered$\}$.

Relation (A.11) appears to be nonidentical. To make it sure that this is true
one must inspect the commutator of the form $\omega $.

Nonidentity of the evolutionary relation points to the nonequilibrium state 
and the development of the gas dynamic instability.
Since the nonequilibrium state is produced by internal forces that are described
by the commutator of the form $\omega $, it becomes evident that the cause
of the gas dynamic instability is something that contributes into the
commutator of the form $\omega $.

One can see (see (A.8)) that the development of instability is caused by
not a simply connectedness of the flow domain,  nonpotential  external
(for each local domain of the gas dynamic system) forces, a nonstationarity
of the flow.

\{In the case when gas is 
nonideal equation (A.9) can be written in the form
$${{\partial s}\over {\partial \xi ^1}} \,=\,A_1 \eqno$$
where $A_1$ is the expression that depends on the energetic actions (transport
phenomena: viscous, heat-conductive).
In the case of reacting gas the extra terms connected with the chemical
nonequilibrium state are added. These factors contributes to the
commutator of the form $\omega $.\}

All these factors lead to emergency of internal forces,
that is, to nonequilibrium state and to development
of various types of instability.

And yet for every
type of instability one can find the appropriate term giving contribution
to the evolutionary form commutator, which is responsible for this type
of instability.
Thus, there is an unambiguous connection between the type of instability
and the terms that contribute to  the evolutionary form commutator in the
evolutionary relation. \{In the general case one has to consider the
evolutionary relations that correspond to the balance conservation laws
for angular momentum and mass as well\}.

As it was shown above, under realization of additional degrees of freedom
it can take place the transition from the nonequilibrium state to the locally
equilibrium one, and this process is accompanied by emergency of physical
structures.
The gas dynamic formations that correspond to these physical structures are
shocks, shock waves, turbulent pulsations and so on. Additional degrees of
freedom are realized as the condition of the degenerate transformation, namely,
vanishing of determinants, Jacobians of transformations, etc. These conditions
specify the integral surfaces (pseudostructures):
the characteristics (the determinant of coefficients at the normal derivatives
vanishes), the singular points (Jacobian is equal to zero), the envelopes
of characteristics of the Euler equations and so on. Under crossing 
throughout the integral surfaces
the gas dynamic functions or their derivatives undergo the breaks.

\subsection*{Electromagnetic field}

The system of charged particles is a material medium, which
generates  electromagnetic field.

If to use the Lorentz force ${\bf F\,= \,\rho (E + [U\times H]}/c)$,
the local variation of energy and linear momentum of the charged
matter (material system) can be written respectively as [14]: $\rho ({\bf U\cdot E})$,
$\rho ({\bf E+[U\times H]}/c)$. Here $\rho$ is the charge
density, ${\bf U}$ is the velocity of charged matter. These
variations of energy and linear momentum are caused by energetic and
force actions and are equal to values of these actions. If to denote
these actions by $Q^e$, ${\bf Q}^i$, the balance conservation laws
can be written as follows:
$$\rho \,({\bf U\cdot E})\,=\,Q^e\eqno(A.12)$$
$$\rho \,({\bf E\,+\,[U\times H]}/c)\,=\, {\bf Q}^i \eqno(A.13)$$

After eliminating  the characteristics of material system (the charged
matter) $\rho$ and ${\bf U}$ by using the Maxwell-Lorentz equations
[14], the left-hand sides of equations (A.12), (A.13) can be expressed only
in terms of the strengths of electromagnetic field, and then one can write
equations (A.12), (A.13) as
$$c\,\hbox{div} {\bf S}\,=\,-{{\partial}\over {\partial t}}\,I\,+\,Q^e\eqno(A.14)$$
$${1\over c}\,{{\partial }\over {\partial t}}\,{\bf S}\,=
\,{\bf G}\,+\,{\bf Q^i}\eqno(A.15)$$
where ${\bf S=[E\times H]}$ is the Pointing vector, $I=(E^2+H^2)/c$,
${\bf G}={\bf E}\,\hbox {div}{\bf E}+\hbox{grad}({\bf E\cdot E})-
({\bf E}\cdot \hbox {grad}){\bf E}+\hbox {grad}({\bf H\cdot H})-({\bf H}\cdot\hbox{grad}){\bf H}$.

Equation (A.14) is widely used while describing electromagnetic
field and calculating  energy and the Pointing vector. But equation (A.15)
does not commonly be taken into account. Actually, the Pointing vector
${\bf S}$ must obey two equations that can be convoluted into the
{\it relation}
$$d\bf S=\,\omega ^2\eqno(A.16)$$
Here $d\bf S$ is the state differential being 2-form and the coefficients
of the form $\omega ^2$ (the second degree form) are the right-hand sides
of equations (A.14) and (A.15).
It is just the evolutionary relation for the system of charged particles that
generate electromagnetic field.

By analyzing the coefficients of the form $\omega ^2$ (obtained from equations
(A.14) and (A.15), one can assure oneself that the form commutator is nonzero.
This means that from relation (A.16) the Pointing vector cannot be found.
This points to the fact that there is no such a measurable quantity
(a potential).

Under what conditions
can the Pointing vector be formed as a measurable quantity?

Let us choose the local coordinates $l_k$ in such a way that one direction
$l_1$ coincides with the direction of the vector ${\bf S}$. Because this
chosen direction coincides with the direction of the vector
${\bf S=[E\times H]}$ and hence is normal to the vectors
${\bf E}$ and ${\bf H}$,
one obtains that $\hbox{div} {\bf S}\,=\,\partial s/\partial l_1$,
where $S$ is a module of ${\bf S}$. In addition,
the projection of the vector ${\bf G}$ on the chosen direction turns out to be
equal to $-\partial I/\partial l_1$.
As a result, after separating from vector equation (A.15) its projection
on the chosen direction equations (A.14) and (A.15) can be written as
$${{\partial S}\over {\partial l_1}}\,=\,-{1\over c}{{\partial I}\over {\partial t}}\,+\,
{1\over c}Q^e \eqno(A.17)$$
$${{\partial S}\over {\partial t}}\,=\,-c\,{{\partial I}\over {\partial l_1}}\,+\,c{\bf Q}'^i\eqno(A.18)$$
$$0\,=\,-{\bf G}''\,-\,c{\bf Q}''^i$$
Here the prime relates to the direction $l_1$, double primes relate to
the other directions. Under the condition $d l_1/d t\,=\,c$ from
equations (A.17) and (A.18) it is possible to obtain the relation in differential
forms
$${{\partial S}\over {\partial l_1}}\,dl_1\,+\,{{\partial S}\over {\partial t}}\,dt\,=\,
-\left( {{\partial I}\over {\partial l_1}}\,dl_1\,+\,{{\partial I}\over {\partial t}}\,dt\right )\,+\,
(Q^e\,dt\,+\,{\bf Q}'^i\,dl_1)\eqno(A.19)$$
Because the expression within the second braces in the right-hand side is
not a differential (the energetic and force
actions have different nature and cannot be conjugated), one can obtain
a closed form only if this term vanishes:
$$(Q^e\,dt\,+\,{\bf Q}'^i\,dl_1)\,=\,0\eqno(A.20)$$
that is possible only discretely (rather than identically).

In this case $dS\,=\,0$, $dI\,=\,0$ and the modulus of the Pointing vector $S$
proves to be a closed form, i.e. a measurable quantity. The integrating
direction (the pseudostructure) will be
$$-\,{{\partial S/\partial t}\over {\partial S/\partial l_1}}\,=\,{{dl_1}\over {dt}}\,=\,c\eqno(A.21)$$
The quantity $I$ is the second dual invariant.

Thus, the constant $c$ entered into the Maxwell equations is defined
as the integrating direction.

1. Cartan E., Les Systemes Differentials Exterieus ef Leurs Application 
Geometriques. -Paris, Hermann, 1945.  

2. Schutz B.~F., Geometrical Methods of Mathematical Physics. Cambrige 
University Press, Cambrige, 1982.

3. Tonnelat M.-A., Les principles de la theorie electromagnetique 
et la relativite. Masson, Paris, 1959.

4. Encyclopedia of Mathematics. -Moscow, Sov.~Encyc., 1979 (in Russian).

5. Bott R., Tu L.~W., Differential Forms in Algebraic Topology. 
Springer, NY, 1982.

6. Wheeler J.~A., Neutrino, Gravitation and Geometry. Bologna, 1960.

7. Haywood R.~W., Equilibrium Thermodynamics. Wiley Inc. 1980.

8. Fock V.~A., Theory of space, time, and gravitation. -Moscow, 
Tech.~Theor.~Lit., 1955 (in Russian).

9. Encyclopedic dictionary of the physical sciences. -Moscow, Sov.~Encyc., 
1984 (in Russian).

10. Dirac P.~A.~M., The Principles of Quantum Mechanics. Clarendon Press, 
Oxford, UK, 1958.

11. Pauli W. Theory of Relativity. Pergamon Press, 1958. 

12. Dafermos C.~M. In "Nonlinear waves". Cornell University Press,
Ithaca-London, 1974. 
      
13. Clark J.~F., Machesney ~M., The Dynamics of Real Gases. Butterworths, 
London, 1964. 

14. Tolman R.~C., Relativity, Thermodynamics, and Cosmology. Clarendon Press, 
Oxford,  UK, 1969.

15. Petrova L.~I., On the problem of development of the flow instability. 
//Basic Problems of Physics of Shock Waves. Chernogolovka, 1987, V.~{\bf 1}, 
P.~1, 304-306 (in Russian). 

16. Prigogine I., Introduction to Thermodynamics of Irreversible 
Processes. --C.Thomas, Springfild, 1955.

17. Rumer Yu.~B. Investigations on 5-Optics. GITTL, Moscow, 1956 (in Russian).

18. Einstein A. The Meaning of Relativity. Princeton, 1953.

19. Weinberg S., Gravitation and Cosmology. Principles and applications of 
the general theory of relativity. Wiley \& Sons, Inc., N-Y, 1972.

20. Dirac P.~A.~M., Proc.~Roy.~Soc., {\bf A133}, 60 (1931).

21. Dirac P.~A.~M., Phys.~Rev., {\bf 74}, 817 (1948).

22. Liepman H.~W., Roshko ~A., Elements of Gas Dynamics. Jonn Wiley, 
New York, 1957

\end{document}